\documentclass[latin1,a4paper]{aa}

\usepackage{amsmath,array}

\usepackage{natbib}
\bibpunct{(}{)}{;}{a}{}{,} 
\usepackage{color}
\usepackage{float}
\usepackage{subfigure}
\usepackage[colorlinks = true,
            linkcolor = blue,
            urlcolor  = blue,
            citecolor = blue,
            anchorcolor = blue]{hyperref}
\usepackage{longtable}
\usepackage{rotating}
\usepackage{lscape}
\usepackage{soul}
\usepackage{multirow}
\usepackage{arydshln}
\usepackage{colortbl}
\usepackage[table]{xcolor}
\usepackage{afterpage}
\usepackage{ftnxtra}
\usepackage{dblfloatfix}
\usepackage{multirow,booktabs}
\usepackage{scalerel}
\usepackage{graphicx}
\usepackage[export]{adjustbox}
\usepackage{listings}
\usepackage{enumitem}
\usepackage{pifont}
\usepackage[toc,page]{appendix}

\usepackage{tabularx,booktabs}
\usepackage{bm,array}
\usepackage{dblfloatfix}
\usepackage{grffile}

\usepackage{cancel}
\usepackage[normalem]{ulem}
\newcommand\redsout{\bgroup\markoverwith{\textcolor{red}{\rule[0.5ex]{2pt}{1pt}}}\ULon}

\definecolor{r}{rgb}{1,0,0}

\DeclareRobustCommand{\ion}[2]{%
\relax\ifmmode
\ifx\testbx\f@series
{\mathbf{#1\,\mathsc{#2}}}\else
{\mathrm{#1\,\mathsc{#2}}}\fi
\else\textup{#1\,{\mdseries\textsc{#2}}}%
\fi}

\newcommand{\lyat}{Ly$\alpha$}
\newcommand{\lya}{Ly$\alpha$\;} 
\newcommand{\msun}{\ifmmode M_{\odot} \else M$_{\odot}$\fi}
\newcommand{\msunyr}{\ifmmode M_{\odot}.{\rm yr}^{-1} \else
M$_{\odot}$ .yr$^{-1}$\fi}
\newcommand{\kms}{\ifmmode {\rm km \; s}^{-1} \else km s$^{-1}$\fi}
\newcommand{\cmtwo}{\ifmmode {\rm cm}^{2} \else cm$^{2}$\fi}
\newcommand{\ergs}{\ifmmode {\rm erg s}^{-1} \else erg s$^{-1}$\fi}
\newcommand{\ergscm}{\ifmmode {\rm erg s}^{-1}{\rm cm}^{-2} \else erg s$^{-1}$ cm$^{-2}$\fi}
\newcommand{\sqarcm}{\ifmmode {\rm arcmin}^{2} \else arcmin$^{2} \:$\fi}

\newcommand{\hh}{\ifmmode {\textrm{H}} \else H\fi}

\newcommand{\hit}{{\ion{H}{i}}}

\newcommand{\mgiit}{{\ion{Mg}{ii}}}
\newcommand{\siiit}{{\ion{Si}{ii}}}
\newcommand{\feiit}{{\ion{Fe}{ii}}}
\newcommand{\ciit}{{\ion{C}{ii}}}

\newcommand{\civ}{{\ion{C}{iv}}}
\newcommand{\ovi}{{\ion{O}{vi}}}
\newcommand{\nv}{{\ion{N}{v}}}

\newcommand{\hi}{{\ion{H}{i}} }

\newcommand{\hplus}{{\ion{H}{$^+$}}}

\newcommand{\mgiid}{\ion{Mg}{ii}~$\lambda\lambda 2796, 2803$}
\newcommand{\feiid}{\ion{Fe}{ii}~$\lambda\lambda 2586, 2600$}
\newcommand{\siiid}{\ion{Si}{ii}~$\lambda\lambda 1190, 1193$}
\newcommand{\ciil}{\ion{C}{ii}~$\lambda 1334$}
\newcommand{\siiib}{\ion{Si}{ii}~$\lambda 1260$}
\newcommand{\mgiia}{\ion{Mg}{ii}~$\lambda 2796$}
\newcommand{\mgiib}{\ion{Mg}{ii}~$\lambda 2803$}

\newcommand{\feiia}{\ion{Fe}{ii}~$\lambda 2586$}
\newcommand{\feiib}{\ion{Fe}{ii}~$\lambda 2600$}
\newcommand{\siiia}{\ion{Si}{ii}~$\lambda 1190$}
\newcommand{\siiiaa}{\ion{Si}{ii}~$\lambda 1193$}

\newcommand{\cii}{\ion{C}{ii} }
\newcommand{\mgii}{\ion{Mg}{ii} }
\newcommand{\feii}{\ion{Fe}{ii} }
\newcommand{\siii}{\ion{Si}{ii} }
\newcommand{\mgplus}{{\ion{Mg}{$^+$}}}
\newcommand{\siplus}{{\ion{Si}{$^+$}}}
\newcommand{\feplus}{{\ion{Fe}{$^+$}}}
\newcommand{\cplus}{{\ion{C}{$^+$}}}

\newcommand{\alphav}{\ifmmode {\alpha_{\mbox{\scaleto{{\rm V{}}}{3.5pt}}}} \else $\alpha_{\mbox{\scaleto{{\rm V{}}}{3.5pt}}}$ \fi}
\newcommand{\alphad}{\ifmmode {\alpha_{\mbox{\scaleto{{\rm D{}}}{3.5pt}}}} \else $\alpha_{\mbox{\scaleto{{\rm D{}}}{3.5pt}}}$ \fi}

\newcommand{\alphavt}{\ifmmode {\alpha_{\mbox{\scaleto{{\rm V{}}}{3.5pt}}}} \else $\alpha_{\mbox{\scaleto{{\rm V{}}}{3.5pt}}}$\fi}
\newcommand{\alphadt}{\ifmmode {\alpha_{\mbox{\scaleto{{\rm D{}}}{3.5pt}}}} \else $\alpha_{\mbox{\scaleto{{\rm D{}}}{3.5pt}}}$\fi}

\newcommand{\rascas}{\textsc{rascas}~}
\newcommand{\rascast}{\textsc{rascas}}
\newcommand{\cloudy}{\textsc{cloudy}~}

\newcommand{\taux}{\ifmmode {\tau_{\scaleto{\rm X}{3.9pt}}} \else $\tau_{\scaleto{\rm X}{3.9pt}}$\fi}

\usepackage{stackengine,scalerel}

\begin{document}

 \title{A public grid of radiative transfer simulations for \lya and metal lines in idealised galactic outflows}

    \titlerunning{A public grid of radiative transfer simulations in idealised outflows}
    \authorrunning{T. Garel et al.}
   \author{T. Garel\inst{1,2}, L. Michel-Dansac\inst{2,3}, A. Verhamme\inst{1,2}, V. Mauerhofer\inst{4}, H. Katz\inst{5}, J. Blaizot\inst{2}, F. Leclercq\inst{6} \and G. Salvignol\inst{2}} 

   \institute{Observatoire de Gen\`eve, Universit\'e de Gen\`eve, 
              Chemin Pegasi 51, 1290 Versoix, Switzerland
          \and
             Centre de Recherche Astrophysique de Lyon UMR5574, Univ Lyon1, ENS de Lyon, CNRS, F-69230 Saint-Genis-Laval, France
          \and
              Aix Marseille Univ, CNRS, CNES, LAM, Marseille, France
          \and
             Kapteyn Astronomical Institute, University of Groningen, PO Box 800, 9700 AV Groningen, The Netherlands
          \and
            Sub-department of Astrophysics, University of Oxford, Keble Road, Oxford OX1 3RH, United Kingdom
          \and
             Department of Astronomy, University of Texas at Austin, 2515 Speedway, Austin, TX 78712, USA
            }

   \date{}

\abstract{The vast majority of star-forming galaxies are surrounded by large reservoirs of gas ejected from the interstellar medium. Ultraviolet absorption and emission lines represent powerful diagnostics to constrain the cool phase of these outflows, through resonant transitions of hydrogen and metal ions. The interpretation of these observations is often remarkably difficult as it requires detailed modelling of the propagation of the continuum and emission lines in the gas. To this aim, we present a large public grid of $\approx 20,000$ simulated spectra which includes \hi \lya and five metal transitions associated with \mgiit, \ciit, \siiit, and \feii that is accessible online at \url{https://rascas.univ-lyon1.fr/app/idealised_models_grid/}. The spectra have been computed with the \rascas Monte Carlo radiative transfer code for $5,760$ idealised spherically-symmetric configurations surrounding a central point source emission, and characterised by their column density, Doppler parameter, dust opacity, wind velocity, as well as various density/velocity gradients. Designed to predict and interpret \lya and metal line profiles, our grid exhibits a wide diversity of resonant absorption and emission features, as well as fluorescent lines. We illustrate how it can help better constrain the wind properties by performing a joint modelling of observed \lyat, \ciit, and \siii spectra. Using \cloudy simulations and virial scaling relations, we also show that \lya is expected to be a faithful tracer of the gas at $T\approx 10^4-10^5$ K, even if the medium is highly-ionised. While \ciit{} is found to probe the same range of temperatures as \lyat, other metal lines merely trace cooler phases ($T \approx 10^4$ K). As their gas opacity strongly depends on gas temperature, incident radiation field, metallicity and dust depletion, we caution that optically thin metal lines do not necessarily originate from low \hi column densities and may not accurately probe Lyman continuum leakage.}

\maketitle

\section{Introduction}
\label{sec:intro}
{\let\thefootnote\relax\footnotetext{Email: \href{mailto:thibault.garel@unige.ch}{thibault.garel@unige.ch}}}

The growth of galaxies within dark matter haloes is primarily driven by gas exchanges with their surroundings. While gas accretion provides the raw material necessary to form stars, feedback mechanisms such as supernovae explosions may process material from the interstellar medium (ISM) into the circum-galactic medium (CGM) and/or intergalactic medium \citep[IGM; see][for a review]{naab2017}. In this picture, galactic outflows are thus likely playing a critical role in the regulation of star formation and in the chemical evolution of galaxies and their environment \citep{dekel06,Hopkins2014,Agertz2015, Muratov2017,Mitchell_2020}.

Outflows are commonly probed by blueshifted absorption lines associated with low-ionisation state (LIS) transitions in the ultra-violet spectrum of star-forming (SF) galaxies \citep{shapley03,Zhu2015,Xu2022,Mauerhofer2021}. In parallel, emission lines such as \hi \lya exhibit a wide diversity of spectral shapes which are often interpreted as signatures of resonant radiative transfer in outflowing neutral hydrogen \citep{verh06,Gronke_2017b,Gurung2021,Blaizot2023,Chang2023}. In the multiphase CGM, these lines are usually seen as probes of the cool gas ($T \approx 10^4-10^5$ K) whereas hotter phases are better traced by higher energy transitions \citep[e.g. \civ, \ovi, \nv ;][]{Tumlinson_2017}. 

Interestingly, the observed properties of \lya and metal lines (e.g. velocity shifts) are sometimes correlated, albeit with significant scatter, which supports the idea that they trace similar gas phases and may therefore be used together to constrain the properties of the outflowing material \citep{steidel2010a,Rivera-Thorsen2015,Marchi2019}. In recent years, deep imaging and integral-field spectroscopic observations have enabled the detection of extended \lyat, \mgiit{} and \feiit{} emission around star-forming galaxies, providing further evidence for the presence of large reservoirs of circumgalactic \hi and metals \citep{steidel2011a,Tang_2014,Wisotzki_2016,Finley2017,Burchett2021,Zabl_2021}. In the JWST era, another key challenge is to establish indirect diagnostics that can help identify Lyman-continuum (LyC) leakage from the sources responsible for cosmic reionisation. Given that hydrogen-ionising radiation emitted by galaxies at $z \gtrsim 5$ cannot be detected directly due to the IGM opacity, alternative methods able to probe the \hi content of galaxies have been suggested in order to gain insight into the escape of LyC photons. The \lya line is often seen as one of the best tracers of LyC leakers since its line profile may encode the \hi optical depth along particular sightlines \citep{Verhamme2017}. Still, \lyat-emitter (LAE) statistics may sharply drop at $z \approx 5-7$ \citep[][]{Konno_2014,Kusakabe_2020,Garel2021,jones2023,Goovaerts_2023} owing to the increasingly neutral IGM \citep[although \lya emission may still be detectable at $z\gtrsim 10$ in some cases, see e.g.][]{Bunker2023}. In this context, we may have to rely on other lines that are not affected by the IGM and therefore visible up to the highest redshifts \citep[e.g. \mgiit, \cii or \siiit;][]{Jaskot2014,Henry2018,Chisholm2020,Katz2022}. 
 
Overall, the joint study of \lya and UV metal lines holds great potential for probing the cool phase of the gas in galactic outflows \citep{steidel2010a,barnes2014,Henry2018}. Nevertheless, inferring gas properties from emission and absorption lines remains non-trivial as resonant scattering effects can significantly alter the observed line shapes through Doppler shifts, line infilling, enhanced dust attenuation of the line due to multiple scatterings, or reprocessed flux into fluorescent channels associated with fine structure level transitions. In this context, it is very useful to interpret observations using numerical simulations that are able to account for these effects. 
Many Monte-Carlo codes have been developed over the years to perform radiative transfer (RT) in idealised configurations and/or as a post-processing step of hydrodynamic simulations \citep[e.g.][]{ahn01,zheng02,Baes2003,cantalupo,dijk06,verh06,laursen2007a,forero-romero2011,orsi2012a,Gronke2014a,Behrens2014,Smith_2015,Michel_Dansac_2020}.

A class of particularly successful idealised models, often referred to as thin shell models, assume a central source surrounded by a spherically-symmetric and uniform layer of gas expanding at constant velocity in which the radiation is propagated before escaping the medium or being absorbed by dust grains \citep[e.g.][]{ahn2003a,verh06,Gronke2015}. Although they are based on simplistic hypothesis, such models provide an efficient way to perform a thorough exploration of the wind parameter space (e.g. velocity, gas column density, temperature, dust opacity, etc) by building large grids of RT simulations \citep[e.g.][]{schaerer11,Gurung2019}. Alternative studies have relaxed some of the thin shell approximations by adding more complexity to the models, such as the inclusion of extended gas outflows (e.g. thick shells), non-spherical geometries, varying wind velocities, or inhomogeneous source/gas distribution \citep{Dijkstra_2012,Zheng_2014,Gronke_2016,Gronke_2017b,Gurung2021}.
Altogether, these various idealised wind models have proven very useful to interpret the \lya line shapes and have shown great success at reproducing the wide diversity of observed spectra \citep{verh08,Hashimoto_2015,Gronke_2017,Orlitova2018}, sometimes in combination with surface brightness profiles \citep{Song2020}. In parallel, similar approaches have been applied to the modelling of prominent UV resonant lines (e.g. \mgiit, \siiit, \feiit) in order to assess the impact of resonant scattering on the observed line shapes through the analysis of infilling, aperture effects or fluorescent decay \citep{Prochaska2011,Scarlata2015,Zhu2015,Carr_2018}.

Building upon previous work, we revisit idealised wind models in order to model simultaneously the spectra of \lya and five prominent UV resonant lines/multiplets associated with \siplus, \cplus, \feplus, and \mgplus. To do so, we generate a large set of RT simulations in spherical winds computed with the \rascas code \citep{Michel_Dansac_2020}. Our models assume either a continuum or a line emission at the centre of a thick expanding shell which is described by six wind parameters, allowing for various gas density and velocity radial profiles. The resulting $\approx 20,000$ spectra have been compiled into a public grid available online and can therefore be used to consistently predict the line profiles of \lya and metallic ions for a given wind configuration.\\[1pt] 

This paper is structured as follows. In Section \ref{sec:lines_set}, we present the UV lines used in this study and review the conditions under which they are expected to trace cool gas phases. Section \ref{sec:simus} describes the radiative transfer simulations and wind models. In Section \ref{sec:integrated_spec}, we examine our grid of simulated spectra by showing the variation of integrated line profiles across the parameter space. Section \ref{sec:appli} presents possible applications of the grid such as the study of photon infilling effects and the joint modelling of \lya and metal lines. In Section \ref{sec:discussion}, we discuss the main assumptions of our wind model and the merits of UV lines as probes of \hi and LyC escape. A summary is given in Section \ref{sec:conclusion}. Appendix \ref{appendix1} presents the online interface to access the grid of simulated spectra.

\section{\lya and metal lines as tracers of cool gas}
\label{sec:lines_set}

\begin{figure*}
\centering
\hskip-2ex
\includegraphics[width=0.99\textwidth]{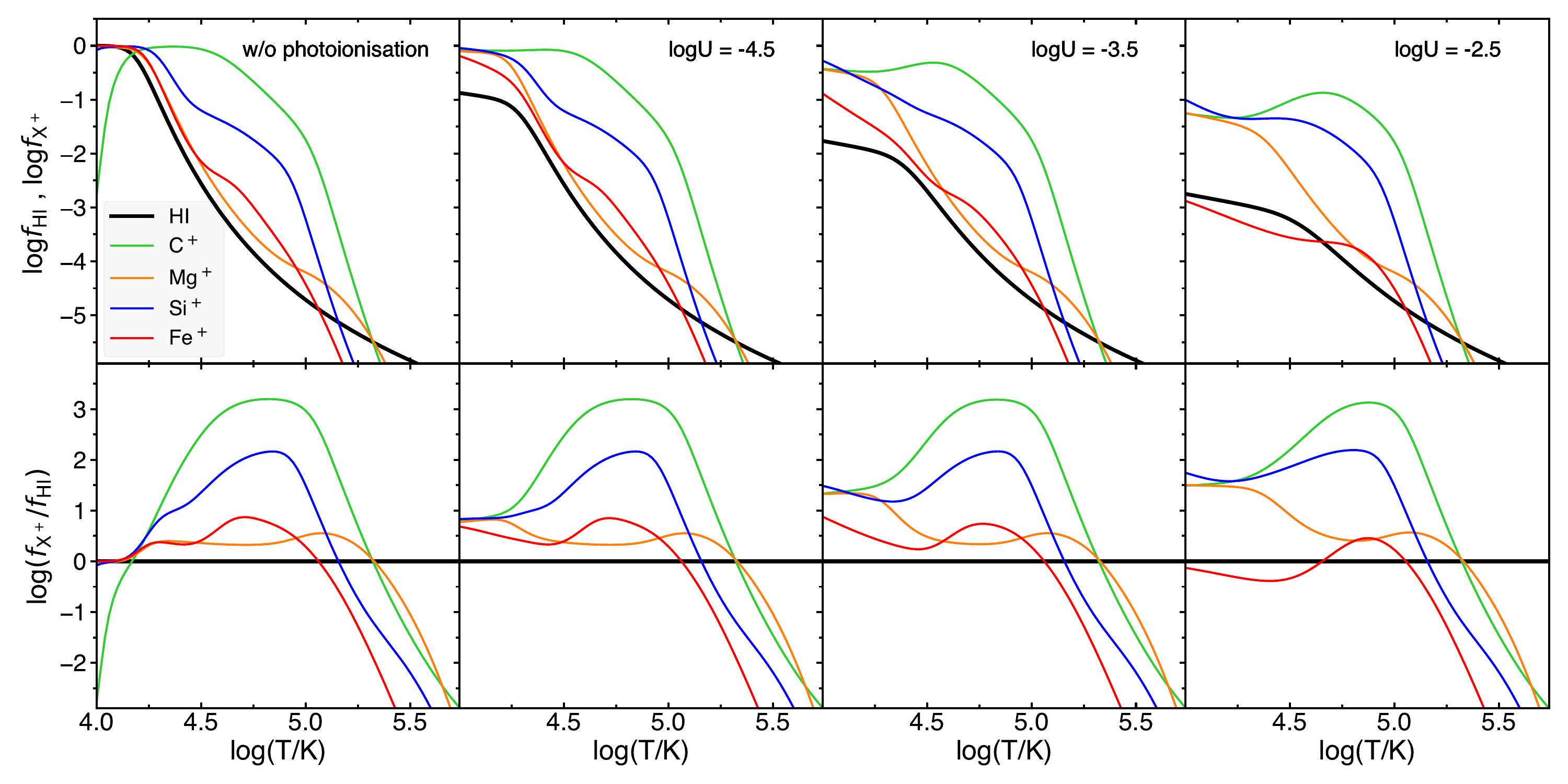}
\caption{\textit{Top:} Fraction of \hit, \mgplus, \siplus, \feplus, and \cplus{} as a function of temperature, assuming collisional ionisation equilibrium (first column) and photoionisation equilibrium (other columns). 
Ionisation fractions have been computed with \cloudy using solar abundance ratios and constant gas density. For photoionisation models, we use incident stellar radiation fields from BPASS 2.2.1 assuming solar metallicity and parametrised by various ionisation parameter values, i.e. log$U=-4.5, -3.5,$ and $-2.5$ (see text for details). \textit{Bottom:} Relative singly-ionised metal fractions with respect to hydrogen neutral fraction.}
\label{fig:ionfrac}
\vskip-2ex
\end{figure*}

In this study, we focus on six sets of resonant UV lines corresponding to transitions observed in the cool gas associated with galaxies and their surrounding medium, namely \hi \lya $\lambda 1216$, the \siiid{} doublet, \siiib, \ciil, the \feiid{} multiplet, and the \mgiid{} doublet. Known as the most intense emission line in the Universe, \lya is mainly powered by recombination and radiative de-excitation following collisional excitation in star-forming regions \citep{Dijkstra2019}. LIS metal lines may similarly arise in singly-ionised phases of the ISM as pure emission in photoionised and collisionally excited/ionised gas, but also through absorption/re-emission of the stellar continuum close to the resonance by the gas \citep[i.e. \textit{continuum pumping};][]{Scarlata2015,Chisholm2020,Katz2022,Xu_2023,Seon23}. As hinted by observed spectra, both \lya and LIS lines are likely often propagated in optically thick gas outflows before escaping galaxies and may thus encode important information about the underlying wind properties \citep{shapley03,Martin_2012,Henry_2015,Rupke2018,Xu2022} . 

The ionisation potential of hydrogen ($E \sim 13.6$ eV) being comprised between the first- and second-ionisation energies ($E \approx 7-11.2$ eV and $E \gtrsim 15$ eV respectively) of Si, C, Fe, and Mg atoms, \hi and singly-ionised metal ions may trace the same media. In order to estimate typical \hit, \siplus, \cplus, \feplus, and \mgplus{} fractions in the cool CGM, we explore different scenarios using the \cloudy code \citep{Ferland2017} and assuming (i) collisional ionisation equilibrium, and (ii) photoionisation equilibrium for various incident radiation intensities. These \cloudy runs are based on one-zone models with constant hydrogen density ($n_{\rm H}=0.001$ cm$^{-3}$) and a gas-phase metallicity fixed to solar values. For photoionisation equilibrium cases, the incident radiation field is given by the spectral energy distribution of a 10 Myr old stellar population at solar metallicity from BPASS 2.2.1 \citep{Stanway_2018} for different values of the dimensionless ionisation parameter $U$, i.e. the ratio of hydrogen-ionising photon density to hydrogen number density. We adopt a Kroupa initial mass function (IMF) with a slope $-1.3$ between 0.1 and 0.5 $M_{\odot}$ and -$2.35$ between 0.5 to 100 $M_{\odot}$ (note that the version 2.2.1 of BPASS includes a fraction of binary stars that depends on the primary star mass). 

From Figure \ref{fig:ionfrac} (top panel), the \hi neutral fraction, $f_{\rm HI}$, and the ionised metal fractions, $f_{\rm X^{+}}$, are both maximal around $T\approx 1-2 \times 10^4$ K, reaching values close to unity when photoionisation is ignored or remains moderate (log$U \lesssim -4$). At higher temperatures or for stronger radiation fields, neutral hydrogen becomes quickly sub-dominant compared to \hplus and $f_{\rm HI}$ drops to values as low as $\approx 10^{-5}$ at $T\approx 10^5$ K. The fractions of singly-ionised metals also tend to decrease at $T > 10^4$ K (owing to the fact that most ions become twice-ionised), though in most cases, they remain higher than $f_{\rm HI}$ over the range $T\approx10^4-10^5$ K (bottom panels of Figure \ref{fig:ionfrac}). Interestingly, the fraction of \cplus{} is predicted to better trace the warm phase compared to the other singly-ionised elements given that $f_{C^+}$ is nearly always greater than $0.1$ at $T \approx 10^4-10^5$ K. When photoionisation is neglected however, $f_{C^+}$ sharply drops at $\lesssim 2 \times 10^4$ K because most carbon elements remain in their neutral form due to the relatively higher ionisation energy of \ion{C} ($E=11.2$ eV) with respect to \ion{Mg}{}\!, \ion{Fe}{} and \ion{Si}{} ($E < 8.2$ eV). In addition, Figure \ref{fig:ionfrac} shows that the fraction of \feplus{} evolves similarly to $f_{\rm HI}$ at $10^4<T<10^5$ K, although it seems to decrease more rapidly for large ionisation parameters (i.e. from log$U=-3.5$ to $-2.5$). In this regime, we find that the $f_{\rm Fe^+}/f_{\rm HI}$ ratio is particularly sensitive to the shape of the input spectrum (i.e. $f_{\rm Fe^+}/f_{\rm HI}$ can change by a factor $2-3$ when varying the age of the stellar population between 5 and 20 Myr), whereas the spectral shape has a much weaker impact on the other ionisation fraction ratios according to our \cloudy simulations.

It is noteworthy that alternative choices regarding the shape of the input UV spectrum (due to e.g. the IMF or the stellar population modelling) and the properties of the gas could impact the derived ionisation fractions. We have checked that our results barely change if we assume different gas densities, subsolar metallicities, different ages of the stellar population (restricting ourselves to ages typical of SF galaxies, i.e. $5-20$ Myr) and if we include the contribution of cosmic microwave background radiation. It remains plausible that other aspects unaccounted for in our study, such as the potential variation of abundance ratios at low metallicity \citep[]{Gutkin_2016} or deviations from our fiducial Kroupa IMF, can further alter the ionisation fractions of hydrogen and metal ions. Nevertheless, these aspects are probably sub-dominant compared to the large variations due to the temperature and the ionising parameter $U$ depicted in Figure \ref{fig:ionfrac}.

\begin{figure}
\hskip-2ex
\includegraphics[width=0.5\textwidth,valign=c,trim=0.cm 5.5cm 1cm 6.8cm, clip]{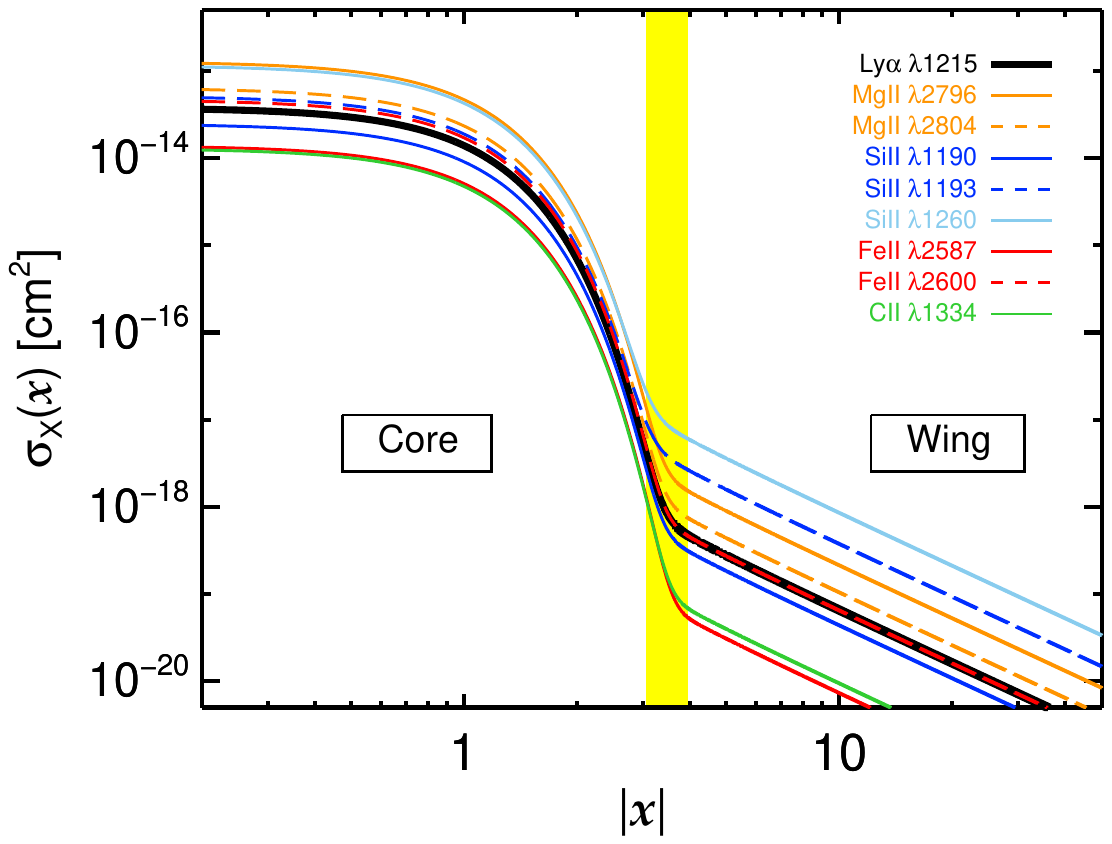}
\vskip-1ex
\caption{Scattering cross-section of \hi \lya $\lambda 1216$, \siiia, \siiiaa, \siiib, \ciil, \feiia, \feiib, \mgiia{} and \mgiib{}. The core-wing transition, highlighted by the yellow stripe, occurs at $|x|\approx3-4$, depending on the lines. The cross-sections have been computed using the atomic parameters listed in Table \ref{table_1} assuming a Doppler parameter value of $b=20$ \kms (see Section \ref{subsec:rt} for a formal definition of these quantities).}
\label{fig:cross}
\end{figure}

Based on these simple \cloudy experiments, \hit, \siplus, \cplus, \feplus, and \mgplus{} are indeed expected to co-exist in the cool CGM. Nevertheless, in order to scatter resonant line photons and leave an observable imprint on galaxy spectra, the gas must be optically thick to these lines ($\tau > 1$). Assuming a homogeneous medium, the total opacity is defined as the product of the cross-section ($\sigma$) and the integrated column density ($N$), namely $\tau = \sigma N$. As highlighted in Figure \ref{fig:cross}, the cross-sections of \lyat, \mgiia, \mgiib, \siiia, \siiiaa, \siiib, \ciil, \feiia, and \feiib{} are quite comparable, at least in the core. The strengths at line centre are of the order of $10^{-13}-10^{-14}$ cm$^2$ and drop by a factor $10^4-10^5$ near the core-wing transition (i.e. $|x| \approx 3-4$, where $x$ is the frequency shift with respect to line centre; see Section \ref{subsec:rt}). Thus, minimum column densities of \hi and metal ions $N \approx 10^{13}-10^{14}$ cm$^{-2}$ are required for the gas to achieve $\tau > 1$. Examining whether or not this condition is usually satisfied for \lya and LIS lines will be discussed in more detail in Section \ref{subsec:line_opacities}.

We now turn our interest to the modelling of UV resonant lines in optically thick galactic outflows. As described in the next section, we have performed a large set of RT simulations in the same wind configurations in order to provide a consistent framework to interpret simultaneously \lya and metal line observations.

\section{Grid of simulations}
\label{sec:simus}

In this section, we present our radiative transfer simulations and describe the wind models that we have explored to build our grid of resonant line spectra. 

\subsection{Emission and radiative transfer}
\label{subsec:rt}

We use the \rascas 3D Monte-Carlo code \citep{Michel_Dansac_2020} to compute numerically the emission and propagation of resonant photons. The detailed properties of the lines considered in this study, namely \hi \lya $\lambda 1216$, the \siiid{} doublet, \siiib, \ciil, the \feiid{} multiplet, and the \mgiid{} doublet, are given in Table \ref{table_1}. Among these, the resonant \siiid, \siiib, \ciil, \feiid{} transitions are coupled with non-resonant decay channels due to fine-structure splitting of their ground state that can produce a pure emission feature, shifted away from the resonance. These \textit{fluorescent} lines are labelled with an asterisk for the remainder of this paper (e.g. C\textsc{ii}$^{*}$ $\lambda 1335$). 

Our simulations model a point source emission with a spectral range encompassing all resonant and fluorescent features of interest from $\lambda_{\rm min}$ to $\lambda_{\rm max}$ (see Table \ref{table_1}). For \siiia, \siiiaa, \siiib, \ciil, \feiia, \feiib, the intrinsic emission has only been modelled as a flat spectrum. In this case, our simulations consist of describing the reprocessed radiation from a stellar-like continuum through resonant pumping and, when relevant, the fluorescent decays\footnote{All species are assumed to be in the ground state throughout this study, owing to the extremely short de-excitation timescales involved (as inferred from Einstein coefficients in Table \ref{table_1}). Hence we neglect the possibility that excited levels can be populated (through e.g. collisional excitation) and give rise to in-situ fluorescent emission \citep[e.g.][]{Xu_2023}.}. \lya and \mgiid{} are however predicted to power strong emission lines in star-forming regions \citep[e.g.][]{Dijkstra_2017,Katz2022}. Therefore, we have also modelled the intrinsic \lya (\mgiit) emission as a flat continuum plus a single (double) Gaussian centred on $\lambda_{\rm 0}=1215.67$ \AA{} ($\lambda_{\rm 0,a}=2796.35$ \AA{} and $\lambda_{\rm 0,b}=2803.53$ \AA). Based on typical recombination line values, we take an intrinsic full width half-maximum of FWHM$_{\rm in}=150$ \kms{} for \lya and both the \mgiid{} doublet lines \citep{Orlitova2018}. We assume a fiducial equivalent width (EW) of 100 \AA{} as an input value for \lya \citep{charlot93}. For the intrinsic \mgiid{} doublet line ratio, we follow \citet{Chisholm2020} and \citet{Katz2022} who show that the intrinsic \mgii emission in SF galaxies predominantly occurs through collisional de-excitation, which sets this ratio to a value of 2 -- corresponding to the ratio of their quantum degeneracy factors $(2J + 1)$. Based on the observational measurements of \citet{Henry2018}, we assume an EW of 6 \AA{} and 3 \AA{} for the \mgiia{} and \mgiib{} lines respectively.

In all cases, photon packets are cast isotropically from a central source and then propagated through the gaseous medium using a Monte-Carlo method. In \rascast, the total optical depth $\tau_{\rm tot}$ through a mixture of gas and dust can be written as:
\begin{equation}
\label{eq:tautot}
\tau_{\rm tot}(\nu) = \sum_{lu}^{\rm transitions} \tau_{X,lu} (\nu) +
\tau_{\rm dust}
\end{equation}
where $\nu$ is the photon frequency in the frame of the gas (and dust)
mixture, $\tau_{X,lu}$ is the gas opacity of a given species $X$ between energy levels $l$ and $u$, and $\tau_{\rm dust}$ is a free parameter representing the integrated dust opacity (see Section \ref{subsec:models}). The sum over transitions is relevant for the cases of line doublets/multiplets for which at least two resonant transitions are modelled simultaneously in our simulations (e.g. \siiid).   
In Eq. \ref{eq:tautot}, $\tau_{X,lu}$ corresponds to the integral along the path of the cross-section $\sigma_{X,lu}(a,\nu)$ of the transition $lu$ times the gas density $n_{X,l}$ : 
\begin{equation}
\tau_{X,lu}(\nu) = \int n_{X,l} \sigma_{X,lu}(a,\nu)  dr
\label{eq:opacity}
\end{equation}

The scattering cross-section is defined as $\sigma_{X,lu}(a,\nu) = \frac{\sqrt{\pi} e^2 f_{lu}}{m_e c\Delta\nu_D} H(a,x)$ where $e$ and $m_e$ are the electron's charge and mass respectively, $c$ is the speed of light, and $f_{lu}$ is the oscillator strength of the transition $lu$. The Doppler width $\Delta \nu_{D}$  is defined as $(b/c)\nu_{lu}$ where $b$ is the mean thermal/turbulent velocity of scatterers that we treat as a free parameter (see Section \ref{subsec:models}). The Voigt parameter is $a = A_{ul} / (4\pi\Delta\nu_D)$, where $A_{ul}$ is the Einstein coefficient for spontaneous emission. The Hjerting-Voigt function $H(a,x)$ describes the convolution of the single-atom Lorentzian cross section with a Gaussian velocity distribution. The dimensionless frequency variable $x = (\nu -\nu_{lu}) / \Delta\nu_D$ represents the frequency shift with respect to resonance in units of Doppler width. It can also be expressed in terms of the velocity offset $V$ as $x = -V/b$.

The interactions of photon packets with dust grains and scatterers of species $X$ are computed through the probabilities $P_{\rm dust} = \tau_{\rm dust} / \tau_{\rm tot}$ and $P_{X,lu} = \tau_{X,lu}/\tau_{\rm tot}$. In the case of dust interaction, the absorption and scattering probabilities are set by the frequency-dependent albedo $a_{dust}$ (see Table \ref{table_1}). We use the phase function of \citet{Henyey_1941} in case of dust scattering \citep[see Section 3.2.2 of][]{Michel_Dansac_2020}. For resonant interactions allowing more than one decay channel, we compute the probability of fluorescent re-emission as the ratio of the Einstein coefficient associated to the downward transition from the upper level $u$ to a lower sub-level ${l_i}$ over the sum of all sub-levels accessible from level $u$, i.e. $A_{ul_i} / \sum_{i} A_{ul_i}$ \citep[see Section 3 in][for more details]{Michel_Dansac_2020}.

\begin{table*}
  \caption{Atomic data for the UV transitions (taken from the NIST database; \texttt{https://www.nist.gov/pml/atomic-spectra-database/}). Columns from left to right : notation of each transition, rest-frame wavelength of the resonant or fluorescent (labelled with an asterisk) transition, Einstein coefficient $A_{ul}$, oscillator strength $f_{lu}$, cross-section at line centre (assuming a Doppler parameter of $b=20$ \kms; see Section \ref{subsec:rt}), transition's lower level, transition's upper level, wavelength range used for the RT simulation of each transition ($\lambda_{\rm min}$ and $\lambda_{\rm max}$), dust albedo $a_{dust}$ \citep{Li_2001}, and gas-phase logarithmic solar abundances \citep[following][ log$\epsilon_{\rm X} =$ log$(X/H)_{\odot} + 12$ where $(X/H)_{\odot}$ is the relative solar number densities of species $X$ and hydrogen]{Asplund2009}. The last column lists for all species the number fraction of atoms that are depleted onto dust, $\delta_{\rm X,0}$, based on MW measurements by \citet{Roman_Duval_2022}.}
\label{table_1}      
\centering                          
\begin{tabular}{p{0.1\textwidth}>{\centering}p{0.1\textwidth}>{\centering}p{0.1\textwidth}>{\centering}p{0.06\textwidth}>{\centering}p{0.11\textwidth}>{\centering}p{0.14\textwidth}>{\centering}p{0.14\textwidth}>{\centering}p{0.05\textwidth}>{\centering}p{0.05\textwidth}>{\centering}p{0.05\textwidth}>{\centering}p{0.05\textwidth}>{\centering\arraybackslash}p{0.05\textwidth}}
  \hline\hline                 
  \noalign{\smallskip}
 Transition & Wavelength & $A_{ul}$ & $f_{lu}$ & $\sigma_{\rm X,0}$ & lower level  & upper level  & $\lambda_{\rm min}$ & $\lambda_{\rm max} $ & $a_{dust}$ & log$\epsilon_{\rm X}$ & $\delta_{\rm X,0}$  \\   
     & (\AA) & (s$^{-1}$) & & ($10^{-13}$ cm$^2$) & & & (\AA) & (\AA) &  & & \\
       \noalign{\smallskip}
\hline                         
H\textsc{i} Ly$\alpha$ & $1215.67$ & $ 6.265 \times 10^8$ & $0.416$ & $0.38$ & $1s$~~~~~~${}^{2}S$~~~~1/2 & $2p$~~~~~~${}^{2}P^{0}$~~3/2  & 1200 & 1230 & 0.32 & 12.0 & -- \\
\hline 
Si\textsc{ii} $\lambda 1190$             & $1190.42$ & $6.53 \times 10^8$ & $0.277$ & $0.25$ & $3s^23p$~~${}^{2}P^{0}$~~1/2 & $3s3p^2$~~${}^{2}P$~~~3/2 & \multirow{4}{*}{1170} & \multirow{4}{*}{1210} & \multirow{4}{*}{0.32} & \multirow{4}{*}{7.51} & \multirow{4}{*}{0.85} \\
Si\textsc{ii}$^{{ }^{*}}$ $\lambda 1194$    & $1194.50$ & $3.45 \times 10^9$ &  --      & -- &$3s^23p$~~${}^{2}P^{0}$~~3/2 & $3s3p^2$~~${}^{2}P$~~~3/2 &  & &  & &  \\
Si\textsc{ii} $\lambda 1193$             & $1193.28$ & $2.69 \times 10^9$ & $0.575$ & $0.51$ & $3s^23p$~~${}^{2}P^{0}$~~1/2 & $3s3p^2$~~${}^{2}P$~~~1/2 &  &  & &  & \\
Si\textsc{ii}$^{*}$ $\lambda 1197$    & $1197.39$ & $1.40 \times 10^9$ &  --      & -- & $3s^23p$~~${}^{2}P^{0}$~~3/2 & $3s3p^2$~~${}^{2}P$~~~1/2  &  &  & &  & \\
\hline 
Si\textsc{ii} $\lambda 1260$             & $1260.42$ & $2.57 \times 10^9$ & $1.22$  & $1.16$ & $3s^23p$~~${}^{2}P^{0}$~~1/2 & $3s^23d$~~${}^{2}D$~~~3/2  & \multirow{2}{*}{1245} & \multirow{2}{*}{1275} & \multirow{2}{*}{0.32} &  \multirow{2}{*}{7.51} &  \multirow{2}{*}{0.85}  \\
Si\textsc{ii}$^{{ }^{*}}$ $\lambda 1265$    & $1265.02$ & $4.73 \times 10^8$ &  --      & -- & $3s^23p$~~${}^{2}P^{0}$~~3/2 & $3s^23d$~~${}^{2}D$~~~3/2 &  & &  & & \\
\hline 
C\textsc{ii} $\lambda 1334$             & $1334.53$ & $2.42 \times 10^8$ & $0.129$   & $0.13$ & $2s^22p$~~${}^{2}P^{0}$~~1/2 & $2s2p^2$~~${}^{2}D$~~~3/2 & \multirow{2}{*}{1320} & \multirow{2}{*}{1350} & \multirow{2}{*}{0.35} & \multirow{2}{*}{8.43} & \multirow{2}{*}{0.30} \\
C\textsc{ii}$^{*}$ $\lambda 1335$    & $1335.67$ & $4.76 \times 10^7$ &  $0.127$     & -- &$2s^22p$~~${}^{2}P^{0}$~~3/2 & $2s2p^2$~~${}^{6}D^{0}$~~3/2 &  & & & & \\
\hline 
Fe\textsc{ii} $\lambda 2587$             & $2586.65$ & $8.94 \times 10^7$ & $0.072$ & $0.14$ & $3d^64s$~~${}^{6}D$~~~9/2 & $3d^64p$~~${}^{6}D^{0}$~~7/2 & \multirow{5}{*}{2570} & \multirow{5}{*}{2640} & \multirow{5}{*}{0.50} & \multirow{5}{*}{7.50} & \multirow{5}{*}{0.95}  \\
Fe\textsc{ii}$^{*}$ $\lambda 2612$    & $2612.65$ & $1.20 \times 10^8$ &  --      & -- & $3d^64s$~~${}^{6}D$~~~7/2 & $3d^64p$~~${}^{6}D^{0}$~~7/2 &  & &  & &  \\
Fe\textsc{ii}$^{*}$ $\lambda 2632$    & $2632.11$ & $6.29 \times 10^7$ &  --      & -- & $3d^64s$~~${}^{6}D$~~~5/2 & $3d^64p$~~${}^{6}D^{0}$~~7/2 &  & & & &  \\
Fe\textsc{ii} $\lambda 2600$             & $2600.17$ & $2.35 \times 10^8$ &  $0.239$ & $0.47$ & $3d^64s$~~${}^{6}D$~~~9/2 & $3d^64p$~~${}^{6}D^{0}$~~9/2 &  & & &  & \\
Fe\textsc{ii}$^{*}$ $\lambda 2626$    & $2626.45$ & $3.52 \times 10^7$ &  --      & -- & $3d^64s$~~${}^{6}D$~~~7/2 & $3d^64p$~~${}^{6}D^{0}$~~9/2 &  & & &  & \\
\hline 
Mg\textsc{ii} $\lambda 2796$             & $2796.35$ & $2.60 \times 10^8$ & $0.608$ & $1.27$ & $2p^63s$~~ ${}^{2}S$~~~1/2 & $2p^63p$~~${}^{2}P^{0}$~~3/2 & \multirow{2}{*}{2770} & \multirow{2}{*}{2820} & \multirow{2}{*}{0.54}  & \multirow{2}{*}{7.60} & \multirow{2}{*}{0.80} \\                                
Mg\textsc{ii} $\lambda 2804$             & $2803.53$ & $2.57 \times 10^8$ & $0.303$  & $0.64$ & $2p^63s$~~${}^{2}S$~~~1/2 & $2p^63p$~~${}^{2}P^{0}$~~1/2 &  & & & & \\
\hline                                   
\end{tabular}
\end{table*}

\subsection{Wind models}
\label{subsec:models}

We have run our simulations with a custom version of \rascas \citep{Michel_Dansac_2020} which has been adapted to perform resonant line transfer in idealised configurations. In practice, we use a 3D high-resolution regular cartesian grid of $256^3$ cells which is sufficient to achieve converged results \citep[see e.g.][]{Carr_2023}. We notably checked on a subset of models that our simulated spectra are unchanged when performing simulations at finer resolution. In all RT experiments presented in this paper, we assume a central point source surrounded by a spherically symmetric gaseous medium, i.e. a \textit{thick shell} with or without dust, which extends from an inner radius $R_{\rm in}$ to the outer radius $R_{\rm out} = 40R_{\rm in}$ where $R_{\rm out}=0.48$ in code units, the box size being equal to one.

As extensively discussed in the literature, various astrophysical processes are capable of launching cool outflows, yielding multiphase kinematics and complex geometries \citep{Heckman1990,Li2017,Fielding2022,Faucher_Gigu_re_2023}. Here, we deliberately focus on testing the response of resonant lines to the radiation transfer in simple, spherical, outflow geometries. These idealised winds are similar to those used by \citet{Prochaska2011} who focussed on the \mgii and \feii lines only, whereas we expand the study to \lya and additional LIS lines. The models are characterised by six wind parameters that are thought to be among the main quantities driving the escape of resonant photons from optically thick media \citep[e.g.][]{verh06, Gronke_2017,Song2020,Chang2023}, namely $N$, $\tau_{\rm d}$, \alphadt, $V_{\rm max}$, \alphav and $b$.  

$N$ is the integrated gas column density of the atom or ion of interest from $R_{\rm in}$ to $R_{\rm out}$. Similarly, we define the integrated dust opacity as $\tau_{\rm d}$, assuming that dust is uniformly mixed with gas. Unlike the thin shell models \citep[used in e.g.][]{verh06,schaerer11}, we allow here for both uniform and non-uniform gas density profiles $n(r)$ and velocity profiles $V(r)$ through the \alphad and \alphav parameters respectively. For the gas density, we assume either a uniform distribution ($\alphad=0$) or an isothermal profile ($\alphad=2$) described by : 
\begin{equation}
n(r) = n_{\rm 0} \left(\frac{R_{\rm in}}{r}\right)^{\scaleto{\alpha}{3.5pt}_{\scaleto{{\rm D}}{3pt}}} \: \:  \text{for  \alphad = 0 or 2 }
\end{equation}
where $n_{\rm 0}$ is set by the integrated gas column density parameter, $N$. For dusty models, the dust density distribution is scaled to the gas density profile with a normalisation factor $\kappa$, i.e. $\tau_{\rm d} = \int_{R_{\rm in}}^{R_{\rm out}} \kappa n(r) dr$, corresponding to our assumed non-zero integrated dust opacity values ($\tau_{\rm d} = 0.5$ or $1$). 

\newcolumntype{C}{>{\centering\arraybackslash}p{2em}}
\begin{table}
\caption{List of wind parameter values: maximal wind velocity $V_{\rm max}$, \hi column density $N_{\rm \textsc{HI}}$, metal ion column density $N_{\rm metals}$, Doppler parameter $b$, dust opacities $\tau_{\rm d}$, wind velocity profile \alphavt, and gas density profile \alphadt. See Section \ref{subsec:models} for a detailed description of these parameters.}
\resizebox{0.48\textwidth}{!}{
\begin{tabular}{ c C C C C C C C C}
\hline
\hline
\noalign{\smallskip}
$V_{\rm max}$ (\kms) \: \: &  \mbox{          } $ \: \: \: \: 0$ & $\: \: 20 \: $ &  $\: \: 50 \: \: $ & $\: \: 100 \: \:  \: \: \: \: $ & $\: \: \:  200 \: \: \: \: $ & $\: \: \: 400$ & $\! \!  \! \! \! \! \! \! \! \! \! \! \! \! \! \! \! \! \! \! \! \! \! \! \! \! \! \! \! \! \! \! \! \! \! \! \! \! 750$  \\
log$(N_{\rm \textsc{HI}} / \cmtwo)$  \: \: & \: & \multicolumn{5}{c}{$\! \! \! \! \! \! 15$ \: to \: $21$ \: by steps of 0.5 dex}  \\
log$(N_{\rm metals} / \cmtwo)$ \: \: & \: & \multicolumn{5}{c}{$\! \! \! \! 13.5$ \: to \: $17$ \: by steps of 0.5 dex}  \\
$b$ (\kms) \: & \multicolumn{2}{c}{$20$} & \multicolumn{2}{c}{$\: \: \: \: 80$} & \multicolumn{2}{c}{$140$}  \\
$\tau_{\rm d}$  \: & \multicolumn{2}{c}{$0$}  & \multicolumn{2}{c}{$\: \: \: \: 0.5$}  & \multicolumn{2}{c}{$1$}  \\
\alphav \: & \multicolumn{2}{c}{ \mbox{ } -1} \: \: & \multicolumn{2}{c}{\: \: \: \: 0} \:  \: & \multicolumn{2}{c}{1}  \\
\alphad \: & \multicolumn{3}{c}{\mbox{ } \mbox{ } \mbox{ } 0}  & \multicolumn{3}{c}{$\! \!  \! \! \! \! \! \! \! \! \! \! \! \! \!  \! \!  2$}   \\
\noalign{\smallskip}
\hline
\hline
\end{tabular}}
\label{tab:grid}
\end{table}

$V_{\rm max}$ parameterises the bulk velocity of the outflow. We consider three different radial velocity profiles, namely a constant-velocity outflow expanding at $V_{\rm max}$ ($\alphav = 0$), a linearly increasing speed ($\alphav = 1$) and a linearly decreasing speed ($\alphav = -1$) that we define as follows :
\begin{equation}
V(r) =  V_{\rm max} \left(\frac{r}{R_{\rm out}}\right)^{\scaleto{\alpha}{3.5pt}_{\scaleto{{\rm V}}{3pt}}}  \:  \text{for \alphav = 0 and 1}, 
\end{equation}
and
\begin{equation}
 V(r) =  V_{\rm max} \left(\frac{R_{\rm out} - \: r}{R_{\rm out} - R_{\rm in}}\right) \: \text{labelled as "\alphav = -1" models.}
\end{equation}

 Our choice of discrete values for wind velocities ($V_{\rm max}= 0 - 750$ \kms) and dust opacities ($\tau_{\rm d}=0 - 1$) is mainly motivated by previous successful attempts at modelling and interpreting UV resonant lines with RT simulations \citep[e.g.][]{laursen09,schaerer11,Gurung2021}.
For $V_{\rm max}$, this is further supported by observational measurements of SF galaxies which infer typical cool gas outflow speeds varying from a few tens to $\lesssim 1000$ \kms{} \citep[e.g.][]{steidel2010a,Heckman_2015}. Our grid also covers a wide range of gas column densities which corresponds to the optically thick regime for our UV lines (i.e. $N \gtrsim 10^{14}$ cm$^{-2}$; see Section \ref{sec:lines_set}).
Metals being orders of magnitude less abundant than hydrogen, we emphasise that gas column densities span a different range for \lya (log$(N_{\rm {\textsc{HI}}}/{\rm cm}^{-2}) = [15;21]$) and for metal lines (log$(N_{\rm metal}/{\rm cm}^{-2}) = [13.5;17]$) and that $N_{\rm {\textsc{HI}}}$ and $N_{\rm metal}$ are treated as independent parameters. The adopted slopes of the density and velocity profiles (\alphad and \alphavt) do not correspond to specific physical wind models and should rather be seen as simple explorations of different types of radial scaling relations (see Section \ref{sec:discussion} for a discussion). 
The Doppler parameter $b$ is the gas velocity dispersion within a cell that sets the line broadening of the absorption profile. This quantity is often expressed as the quadratic sum of the turbulent and thermal velocities of the gas. The latter is proportional to $\sqrt{T/m_{\rm X}}$, where $m_{\rm X}$ is the atom/ion mass, and takes values of $\approx 13$ \kms{} for hydrogen and only $\lesssim 2$ \kms{} for the metal elements at $T \approx 10^4$ K. Here, we have assumed that the Doppler parameter is dominated by turbulent motions. Thus, in our grid of models we have considered a typical value of 20 \kms{} as well as more extreme cases, i.e. $b=80$ and 140 \kms{} \citep[e.g.][]{schaerer11,Rudie_2012,Chen_2023,Koplitz_2023}.

In total, we have generated $5,760$ different numerical configurations corresponding to all possible combinations of the physical parameter values of our wind models, as summarised in Table \ref{tab:grid}. This translates into a dataset of $\approx 20,000$ RT experiments for the six different lines, i.e. \hi \lya and five metal ion transitions, which allows us to build a large homogeneous grid of spectra able to self-consistently predict the properties of these lines after propagation in the same medium.  All spectra are compiled into a public grid that can be accessed online through a web interface\footnote{\url{https://rascas.univ-lyon1.fr/app/idealised_models_grid/}}, that we describe in Appendix \ref{appendix1}. 

\section{Predicted line profiles}
\label{sec:integrated_spec}
In this section, we examine \lya and metal line profiles as a function of the wind parameters across our grid. We have assumed a Gaussian$+$continuum intrinsic emission for \lya and \mgii whereas we use a flat continuum as an input for \ciit, \siiit, and \feii simulations. In practice, we vary one parameter at a time and fix the others to the following values : log$(N/{\rm cm}^{-2})=19$ for \lya (log$(N/{\rm cm}^{-2})=14.5$ for metals), $V_{\rm max} = 200$ \kms, $b=20$ \kms, $\tau_{\rm d} = 0$, $\alphav = 1$, and $\alphad = 2$. Consequently, the predictions presented in this section only cover a subset of the models and we emphasise that our grid exhibits a much broader diversity of line shapes than what is shown in the figures.  

\subsection{Column density}
\begin{figure}
\hskip-2ex
\includegraphics[width=0.54\textwidth,valign=c,trim=0.0cm 0.1cm 0 4.cm, clip]{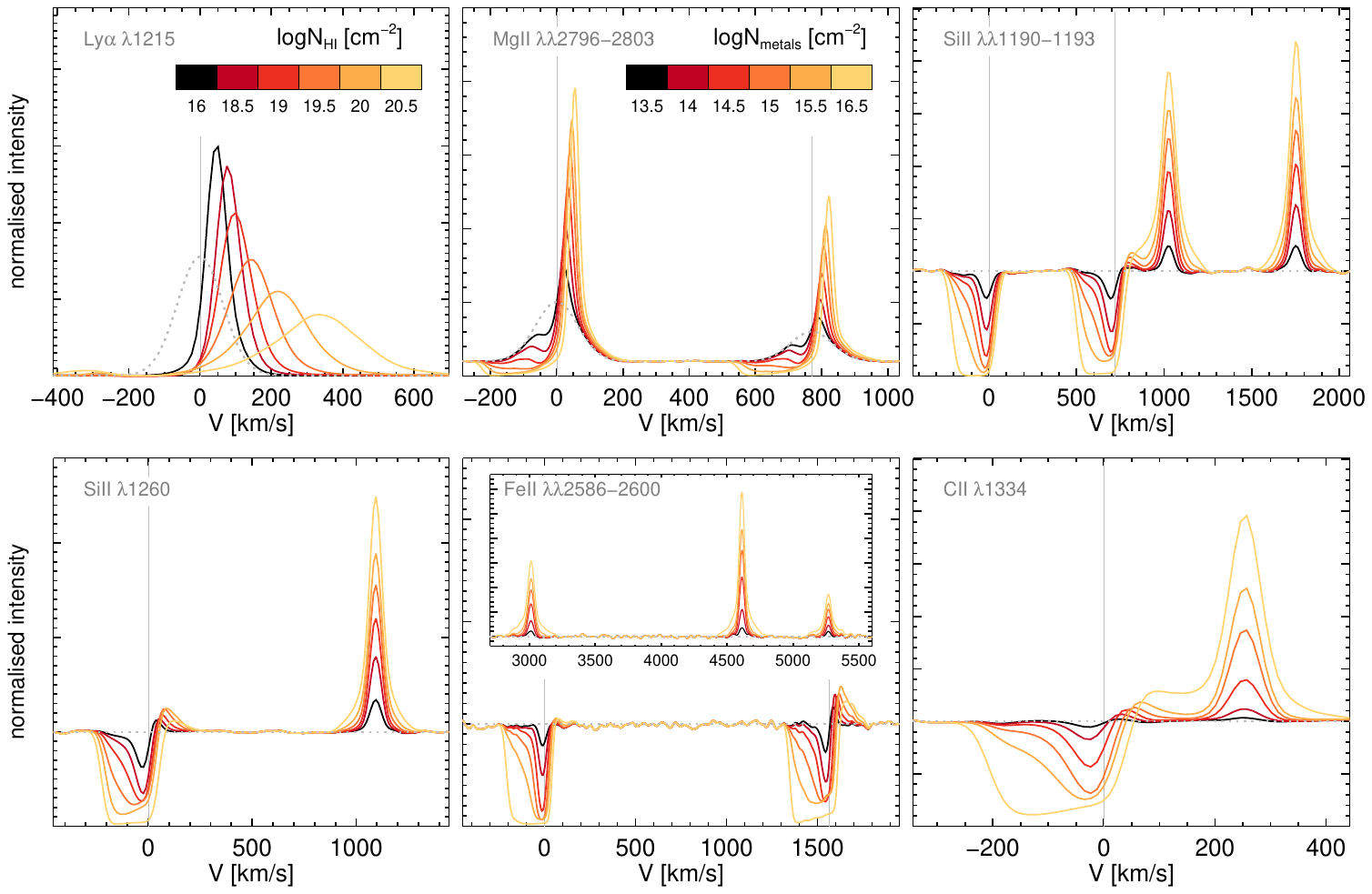}
\vskip-2ex
\caption{Variation of the integrated line profiles as a function of the column density, $N$. Each panel corresponds to one particular line/multiplet, i.e. \hi \lya $\lambda 1216$, \mgiid, \siiid, \siiib, \feiid, and \ciil. All lines have been normalised to the continuum intensity. Dotted grey lines depict the input emission and vertical lines show the location of the resonances. On the x-axis, frequencies are expressed in terms of velocity offset relative to the resonant wavelength $\lambda_{\rm 0}$ (the bluest line is used in case of multiplets), namely $V=c(\dfrac{\lambda}{\lambda_{\rm 0}}-1)$. For \feiit, the inset shows the fluorescent lines associated with \feiid{} which are emitted at much longer wavelengths (see Table \ref{table_1}).}
\vskip-2ex
\label{fig:spec_N}
\end{figure}

In Figure \ref{fig:spec_N}, we study the variation of each line for various values of column density, $N$. When increasing the \hi column density, the red peak of the \lya line becomes broader and more redshifted, with some photons sometimes escaping at $V \gg V_{\rm max}$ for the highest $N_{\rm HI}$ values. This is a well known and generic result of \lya transfer \citep[e.g.][]{laursen09}, which is not only valid for accelerating winds (as shown in Figure \ref{fig:spec_N}) but also for constant or decreasing velocity models. It comes from the increasing number of resonant scatterings associated with expanding gas at higher optical depth, in which blue photons preferentially escape on the red side, sometimes through backscatterings on the receding part of the wind\footnote{A backscattered photon is Doppler boosted by $V(r)cos\theta$ when re-emitted by the receding part of the wind towards the observer; $\theta$ being the angle between the incoming and outgoing directions of the photon ($\theta \in [-\pi/2;\pi/2]$).} \citep[see e.g. Figure 12 of][]{verh06}.
Likewise, outflows with higher \mgplus{} column densities tend to be more opaque to photons emitted blueward of the resonances of the \mgii doublet. Absorbed photons can be redistributed to the red side of the line and then produce highly asymmetric emission lines. When varying $N$, the \mgii emission lines are only moderately broadened and redshifted compared to \lyat, mainly because Mg$^+$ column densities span much lower values than for \hi in our grid.

For lines produced only by continuum pumping (i.e. \ciil, \feiid, \siiid, and \siiib), the expanding gas preferentially absorbs radiation on the blue side of the resonance\footnote{Unlike \lyat, metal line photons are very unlikely to scatter in the wings since most media become highly transparent at $|x \gg 1|$; hence, most interactions are expected to occur close to the core (see Figure \ref{fig:cross}).}. These photons can undergo one or several scatterings in order to escape the medium. This process often yields a typical P-Cygni profile with a blue absorption associated with an emission feature redward of the resonance which is mainly due to strong Doppler shifts acquired by backscattered photons. When the column density increases, the blueshifted absorption becomes more prominent (or even saturated for very high $N$ values) while the strength of the fluorescent lines is boosted. Indeed for larger gas opacities, the number of absorbed photons and the number of scatterings per photon both increase. As seen in Section \ref{subsec:rt}, at each scattering, photons have a fixed probability to decay to another level than the ground state, and then be reprocessed into a fluorescent channel at longer wavelength. Hence, the fluorescent line is amplified when $N$ increases, unless dust attenuation is important (see Section \ref{subsubsec:dust}).

The asymmetric absorption features seen in Figure \ref{fig:spec_N} directly result from the radially-decreasing opacity associated with the wind parameters chosen in this example ($\alphav = 1$ and $\alphad = 2$). Indeed, a photon emitted on the blue side, say at $-V$, is seen close to the resonance by a layer of gas outflowing at $\approx V$. Since the gas density decreases while the wind accelerates towards larger radii, bluer photons are subject to a lower opacity on their way out of the wind compared to photons close to the resonance. This is most visible for low column densities where the absorption extends from $V\approx -V_{\rm max}$ with a dip at $V\approx 0$. For large column densities however, the situation is reversed as the absorption now peaks at $V\approx  -V_{\rm max}$, and an additional emission feature appears close to the resonance at $V \gtrsim 0$. As discussed in \citet[][ e.g. their Figure 5]{Prochaska2011}, this is mainly the result of infilling which redistributes scattered photons around their absorbed frequency: in opaque outflowing media, nearly all photons on the blue side are scattered at least once. Some may then escape the wind and fill-in the line near $V\approx 0$ while others may instead keep scattering until being destroyed by dust, if any, or escape through a fluorescent channel. We will illustrate these effects in more detail in Section \ref{subsec:infill}.

\begin{figure}
\hskip-2ex
\includegraphics[width=0.54\textwidth,valign=c,trim=0.0cm 0.1cm 0 3.8cm, clip]{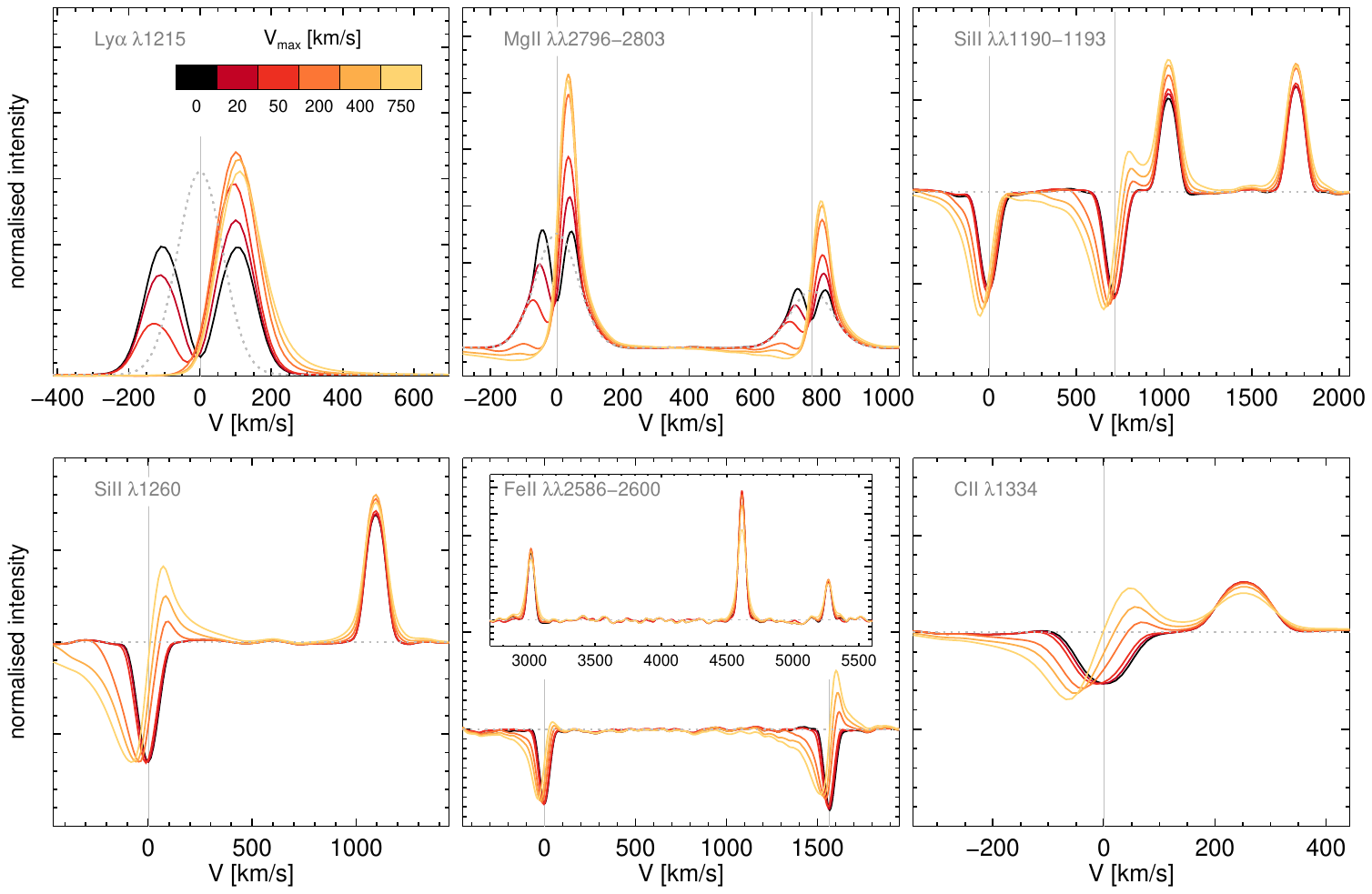}
\vskip-2ex
\caption{Variation of the integrated line profiles as a function of the wind velocity, $V_{\rm max}$. See caption of Figure \ref{fig:spec_N} for a detailed description of the panels.}
\label{fig:spec_vmax}
\end{figure}

\subsection{Wind velocity}
The maximal wind velocity has also a strong impact on each of the lines, as can be seen from Figure \ref{fig:spec_vmax}. For \lya and \mgiit, increasing $V_{\rm max}$ progressively reduces the blue peak relative to the red peak. This is purely because photons on the blue side of the line propagating outwards are seen closer to the resonance (in the frame of the outflowing medium) and are thus less likely to escape directly. 

For the other lines, higher $V_{\rm max}$ values broaden, blueshift and increase the asymmetry of the absorption feature. At $V_{\rm max} \gtrsim 200$ \kms, the effect of infilling becomes stronger such that the absorption centroid is blueshifted. In parallel, more photons gain sufficient Doppler shifts to escape at $V > 0$ on the red side of the line, mainly due to backscatterings. We also note that larger wind velocities usually broaden the fluorescent lines. This effect is not clearly visible in Figure \ref{fig:spec_vmax} because the case considered here assumes $\alphav = 1$ and $\alphad = 2$ which is more optically thin at large radii, hence favouring scatterings with low velocity gas in the inner part of the wind. But in general, fluorescent photons re-emitted by an atom with a radial velocity $V$ will directly escape the medium (unless they interact with dust, if any) at observed frequencies between $-V$ and $+V$ around the fluorescence, depending on the angle between its incoming and escaping directions. In practice, the blue side of the fluorescent emission is dominated by photons re-emitted from the forthcoming part of the wind whereas the red side is rather made of backscattered radiation.  
\begin{figure}
\hskip-2ex
\includegraphics[width=0.54\textwidth,valign=c,trim=0.0cm 0.1cm 0 3.8cm, clip]{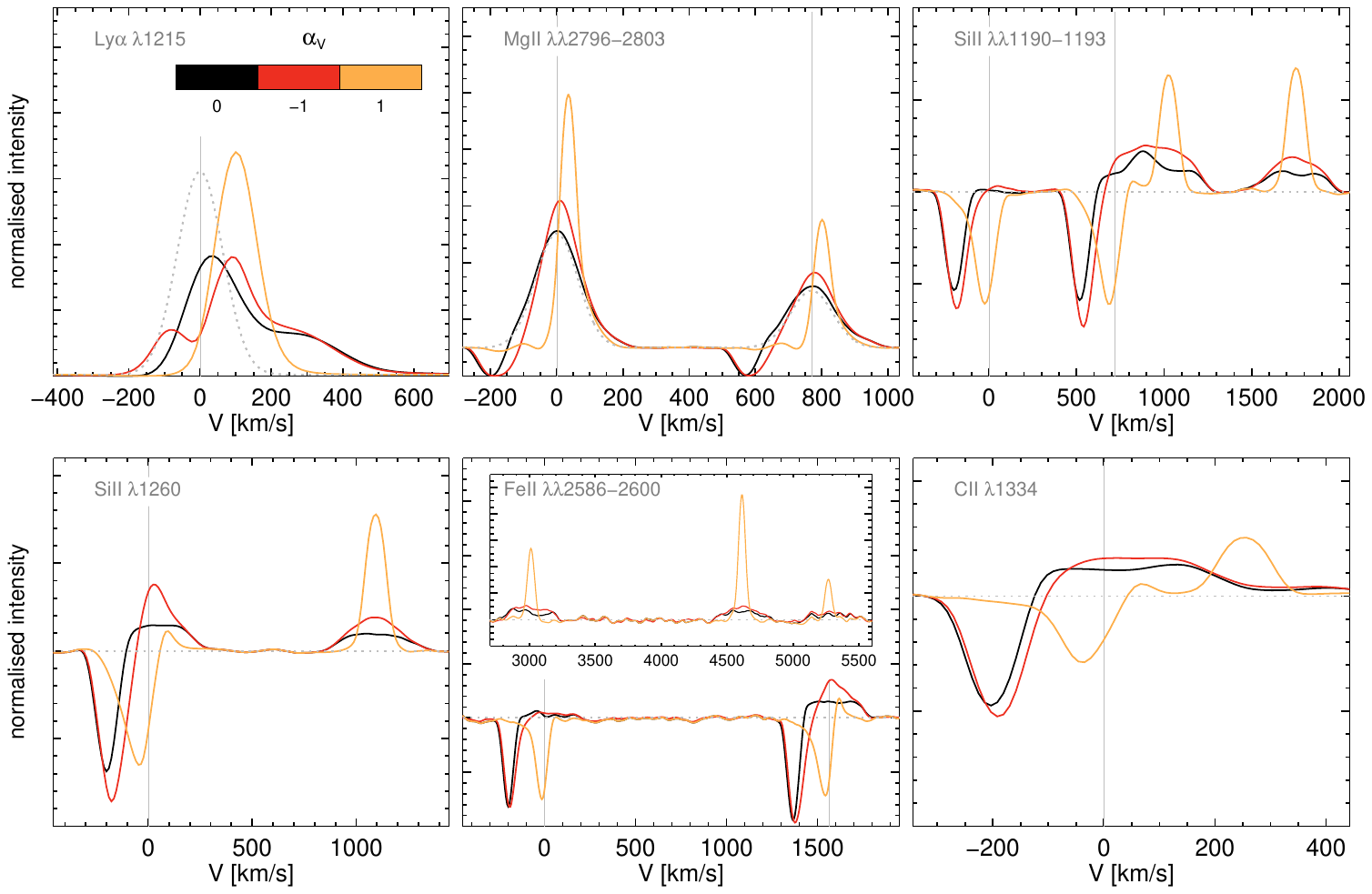}
\vskip-2ex
\caption{Variation of the integrated line profiles as a function of the wind velocity gradient, \alphavt. See caption of Figure \ref{fig:spec_N} for a detailed description of the panels.}
\label{fig:spec_alphav}
\end{figure}
\subsubsection{Velocity gradient}
The radial velocity profile can significantly modify both the emission and absorption line profiles (Figure \ref{fig:spec_alphav}). For \lyat, a noticeable feature is that decelerating models (\alphav$=-1$) often yield double-peak profiles (i.e. a dominant red peak and a smaller blue bump) whatever the value of $V_{\rm max}$. In contrast, double peaks are usually restricted to static media \citep{neufeld1990a} or slow constant-velocity and accelerated winds \citep[e.g. Figure \ref{fig:spec_vmax}; see also Figure 3 in][]{Verhamme2015a}. Although not visible in Figure \ref{fig:spec_alphav}, a similar behaviour often exists for \mgiit{} when the gas column density is high enough.

For metal lines, the peak of absorption is expected at $V=-V_{\rm max}$ for constant velocity models because photons predominantly scatter close to line centre. When the wind velocity varies radially, the absorption traces the location of maximal opacity, respectively at $V\approx 0$ for \alphav$=1$ and at  $V=-V_{\rm max}$ for \alphav$=-1$. We also note that the broadening of the fluorescent line may offer valuable insight into the gas kinematics. For instance, in the \alphav$=0$ case, where all the gas is at $V_{\rm max}$, the fluorescence is broad and extends from $\approx -V_{\rm max}$ (forward scattering) to $\approx V_{\rm max}$ (backscattering). Fluorescent lines are however much narrower for accelerating winds (\alphav$=1$) as photons preferentially have their last scattering before escape at small radii, i.e. within low-velocity gas.

\begin{figure}
\hskip-2ex
\includegraphics[width=0.54\textwidth,valign=c,trim=0.0cm 0.1cm 0 3.9cm, clip]{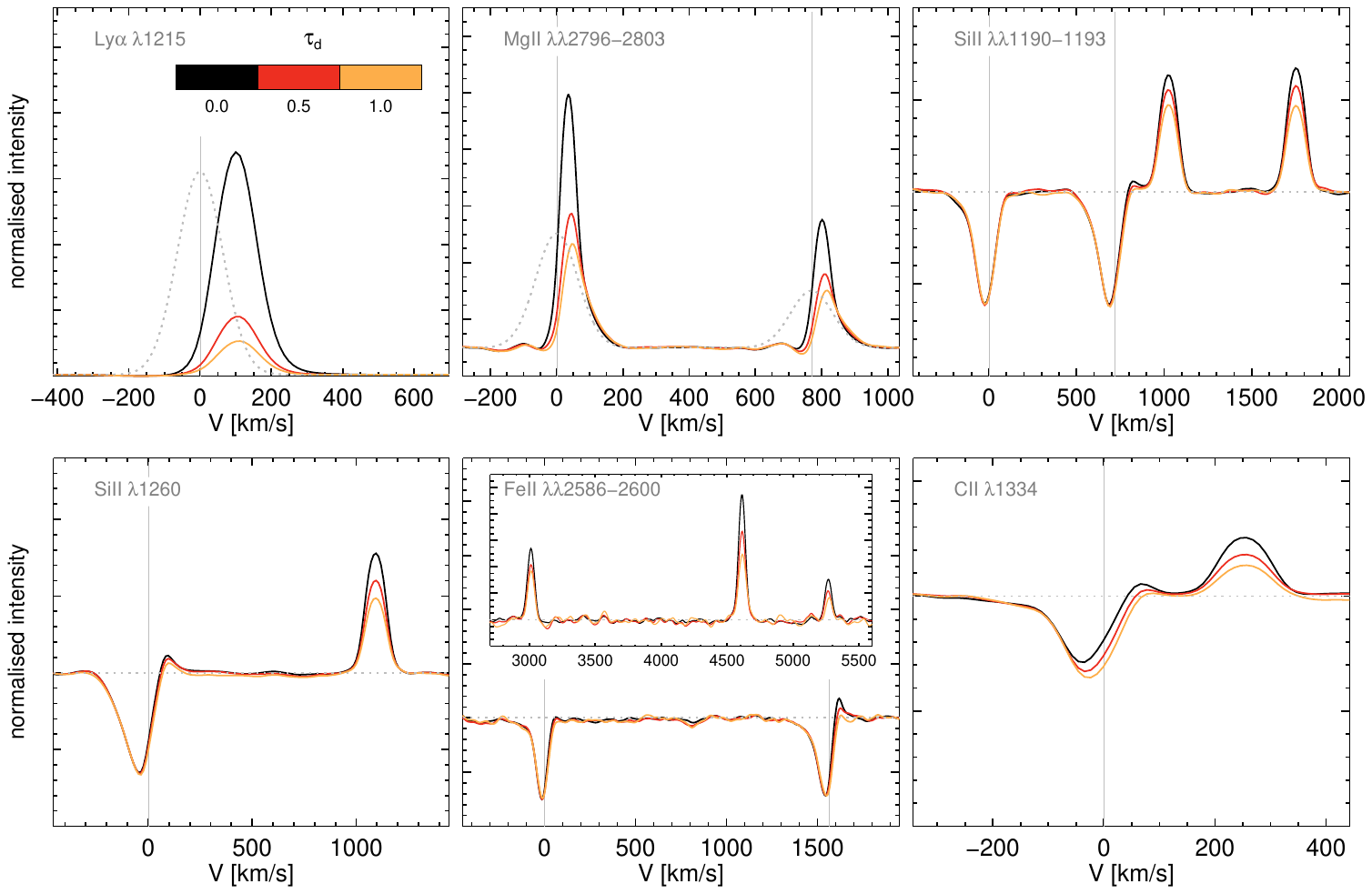}
\vskip-2ex
\caption{Variation of the integrated line profiles as a function of the dust opacity, $\tau_{\rm d}$. See caption of Figure \ref{fig:spec_N} for a detailed description of the panels.}
\label{fig:spec_taud}
\end{figure}

\subsection{Dust opacity}
\label{subsubsec:dust}
In our models, the dust opacity follows the radial distribution of the gas through the \alphad parameter. We highlight that its integrated value, $\tau_{\rm d}$, is independent of the total gas column density parameter, $N$. Therefore, our grid plausibly allows for extreme configurations such as dust-free, high column density, gas or dust-rich gas at very low column densities. In addition, dust grains are assumed to either scatter photons coherently according to a fixed albedo value for each line/multiplet (see Table \ref{table_1}) or absorb them, in which case they are considered to be destroyed. 

As shown in Figure \ref{fig:spec_taud}, the effect of dust on integrated spectra is often rather intuitive because it mostly acts as a destructive term. In optically thick dusty media, multiple resonant scatterings enhance the path traveled by photons which increases the probability of dust absorption. For emission lines like \lya and \mgiit, this has mostly the effect of reducing their strength and equivalent width. For \mgiit, we note that the presence of dust can also affect the flux ratio of the doublet due to the different oscillator strengths of each resonant transition \citep{Katz2022,Seon23}. Regarding lines powered by continuum pumping, both the resonant and fluorescent emissions may be reduced in the presence of dust. First, backscattered photons re-emitted redward of the resonance travel through a large portion of the wind, which can boost dust absorption and then erase the emission feature at $V\gtrsim 0$. Second, while fluorescent lines are produced by successive resonant scatterings (each scattering being associated to a probability of fluorescent decay), resonantly trapped photons are also more likely to be destroyed by dust, which unavoidably decreases the chance of escape through the fluorescence. 

\subsection{Doppler parameter}
\begin{figure}
\hskip-2ex
\includegraphics[width=0.54\textwidth,valign=c,trim=0.0cm 0.1cm 0 3.8cm, clip]{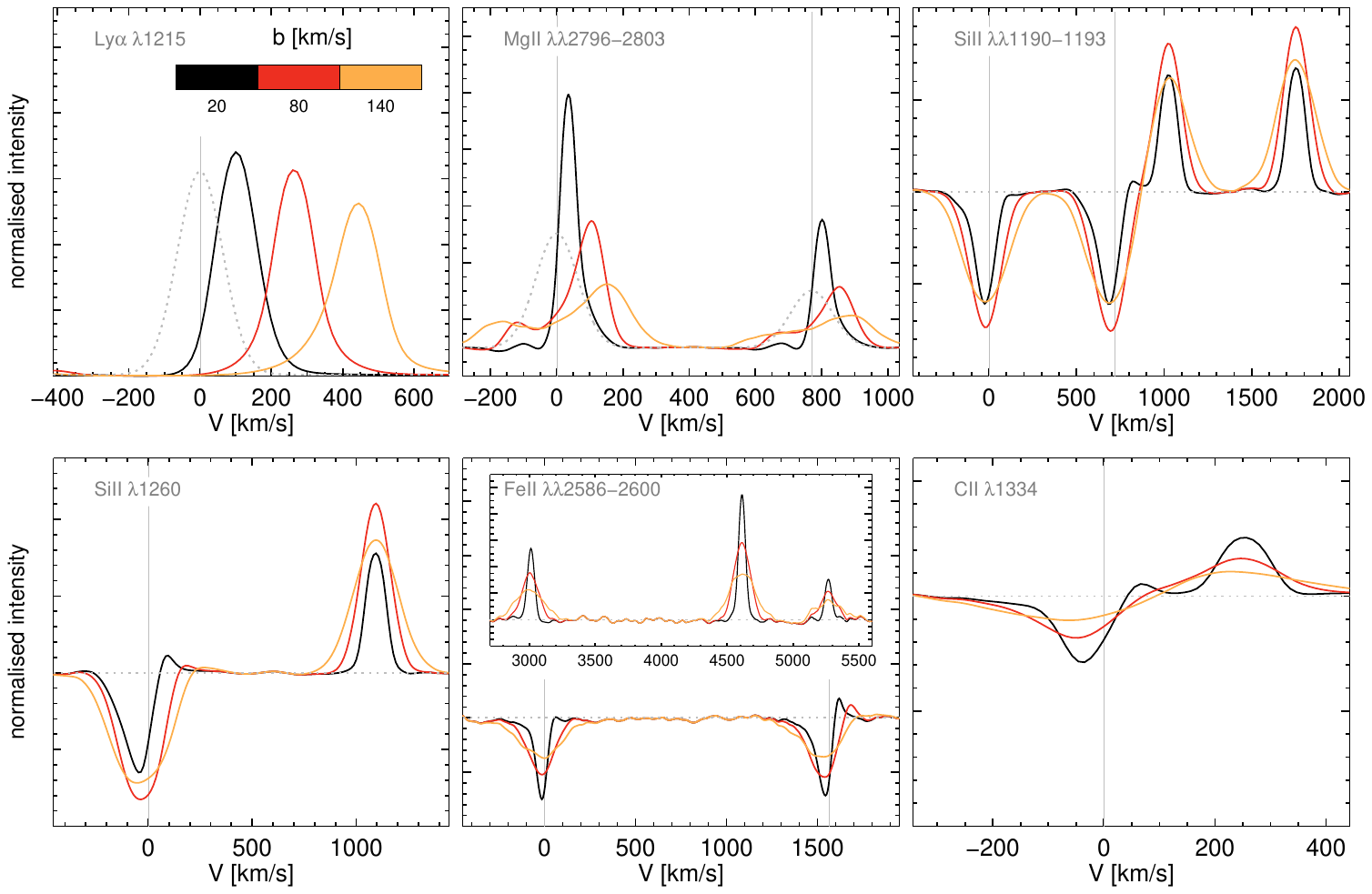}
\vskip-2ex
\caption{Variation of the integrated line profiles as a function of the Doppler parameter, $b$. See caption of Figure \ref{fig:spec_N} for a detailed description of the panels.}
\label{fig:spec_b}
\end{figure}

In addition to the typical Doppler parameter value of $b=20$ \kms{} often assumed in the literature, our grid also explores higher velocity dispersions (80 and 140 \kms) which may be found in highly turbulent media. As expected, using larger $b$ values produces broader metal absorption lines and fluorescent lines, due to the increased probability of scattering by high turbulent-velocity atoms (Figure \ref{fig:spec_b}). For \lyat, the emission line tends to shift to the red as $b$ takes larger values. As mentioned earlier, resonant \lya photons are often trapped in optically thick gas if their frequency is close to the core, so they preferentially escape from winds through wing scatterings. The typical \lya core-to-wing transition, expressed in Doppler units, is of the order of $|x| \gtrsim V/b\approx 3$. Hence, when $b$ is large, photons must redshift to higher velocities to reach the wing and eventually escape. 
\subsection{Density gradient}
\begin{figure}
\hskip-2ex
\includegraphics[width=0.54\textwidth,valign=c,trim=0.0cm 0.1cm 0 3.8cm, clip]{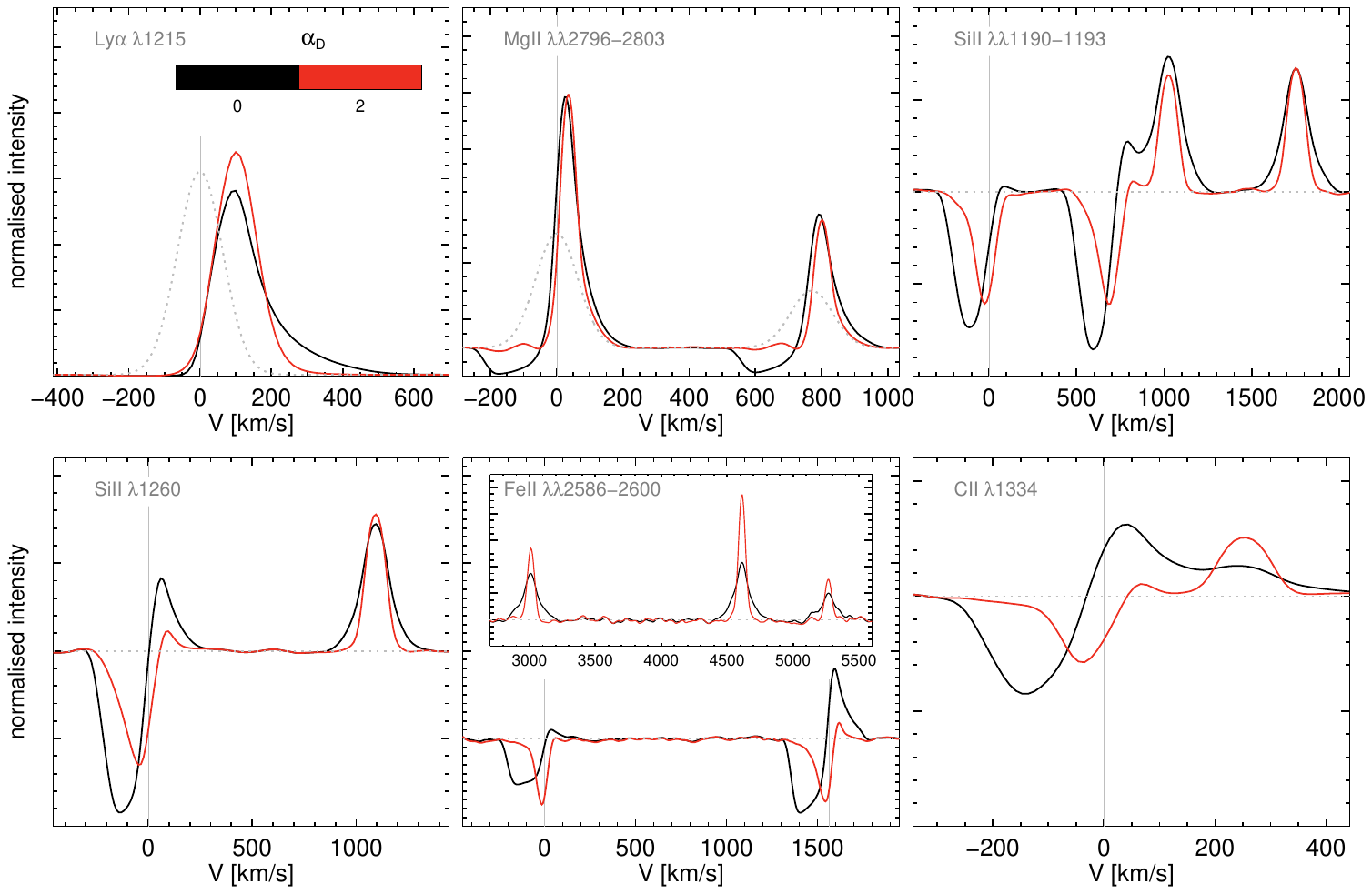}
\vskip-2ex
\caption{Variation of the integrated line profiles as a function of the wind density gradient, \alphadt. See caption of Figure \ref{fig:spec_N} for a detailed description of the panels.}
\label{fig:spec_alphad}
\end{figure}

When varying the gas density profile, \alphadt, the radial opacity seen by propagated photons will strongly depend on the associated velocity profile model, \alphavt. In the example shown in Figure \ref{fig:spec_alphad}, the wind is accelerating radially (\alphav$=1$) so the lines interacting with the gas only through continuum pumping (\siiit, \feiit, \ciit) tend to be mostly absorbed at $V \approx 0$ if most of the gas sits in the central part of the wind (i.e. \alphad$=2$). In the uniform density case (\alphad$=0$), the probability of scattering is constant for photons emitted between $V \approx -V_{\rm max}$ and $V \approx 0$ which significantly broadens the absorption feature. In this case, we note that the absorption centroid can be shifted to $V \approx -V_{\rm max}$ because of line infilling (see Section \ref{subsec:infill}).

Although \alphad can affect the blueshifted absorption of \lya and \mgii in the same way as the \siiit, \feiit, \cii lines, their emission signatures are not strongly altered by the density profile. This is mostly due to the fact that the integrated gas column density remains similar in both cases. Still, the radial opacity in the uniform and isothermal density profiles are completely different by construction such that last scatterings do not occur at the same radii. Therefore, noticeable differences between $\alphad=0$ and $\alphad=2$ models are rather expected in terms of extended emission maps and spatially-resolved spectra (Garel et al., in prep). 

\begin{figure*}
\hspace{-0.4cm}
\vskip-2ex
\includegraphics[width=0.99\textwidth,valign=c,trim=1.cm 2.8cm 0 9.8cm, clip]{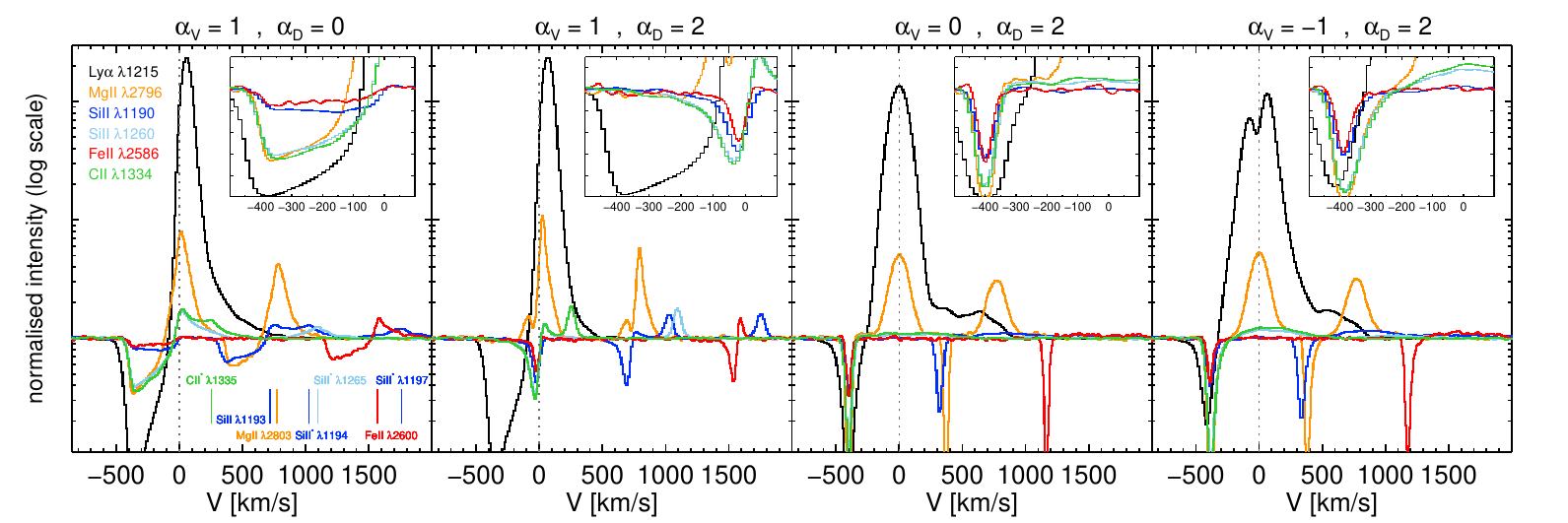}
\caption{Relative variation of \lya and metal line spectra for various values of \alphav and \alphadt. The x-axis shows the frequency in velocity units where the zero point is set at the resonance of the line (for doublets, the line with the shortest wavelength is used) and denoted by a dotted grey line. Note that the normalised intensity in the main panels is in log scale to highlight both emission and absorption features. In each panel, the inset zooms on the absorption region on the blue side of the resonance (y-axis now in linear scale). The values for the fixed wind parameters are $V_{\rm max} = 400$ \kms, $b=20$ \kms, and $\tau_{\rm d} = 0$. For \lyat, the \hi column density is $N_{\rm HI}=10^{18} {\rm cm}^{-2}$. The \mgplus, \feplus, \siplus{} column densities are fixed to $N=10^{14} \: {\rm cm}^{-2}$ whereas we take $N=10^{15} \: {\rm cm}^{-2}$ for \cplus, owing to the fact that carbon is roughly ten times more abundant than the other elements according to solar values.}
\label{fig:allspec}
\vskip-2ex
\end{figure*}

\section{Examples of applications}
\label{sec:appli}

\subsection{Joint modelling of \lya and metal lines}
\label{subsec:jointfit}

A straightforward application of our grid is to predict several lines simultaneously based on the same wind model. Figure \ref{fig:allspec} shows such joint modelling for \lya and metal line spectra assuming different density and velocity profiles. While the $b$, $V_{\rm max}$, $N$, and $\tau_{\rm d}$ parameters are fixed to common values in all panels here, we clearly see that different \alphav and \alphad values strongly alter the spectral lines, owing to the different radial opacities seen by the photons on their way out of the wind. As discussed for each individual line in Section \ref{sec:integrated_spec}, the most noticeable variations are the line shapes of resonant emissions, the position of absorption features, the intensity of the fluorescences, the asymmetry of the absorption, and the degree of saturation. 

\begin{figure}[!h]
\vskip-1ex
\includegraphics[width=0.48\textwidth,valign=c,trim=0.cm 10.1cm 0 7cm, clip]{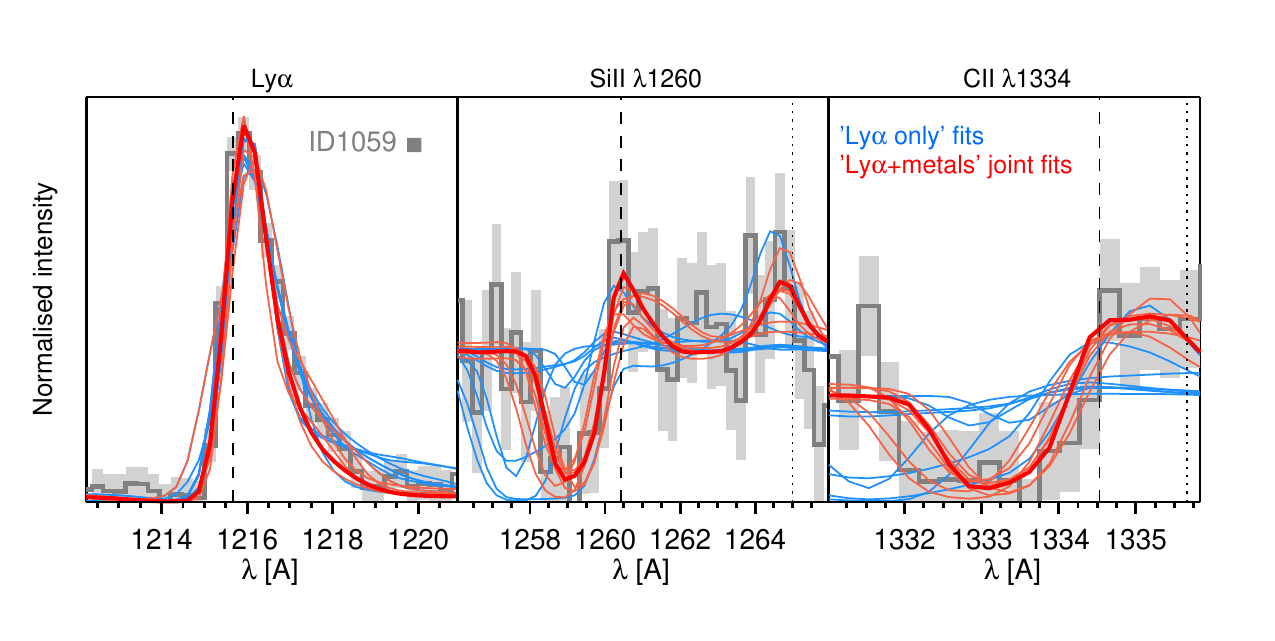}
\includegraphics[width=0.51\textwidth,valign=c,trim=1.5cm 1.2cm 0.7cm 4.cm, clip]{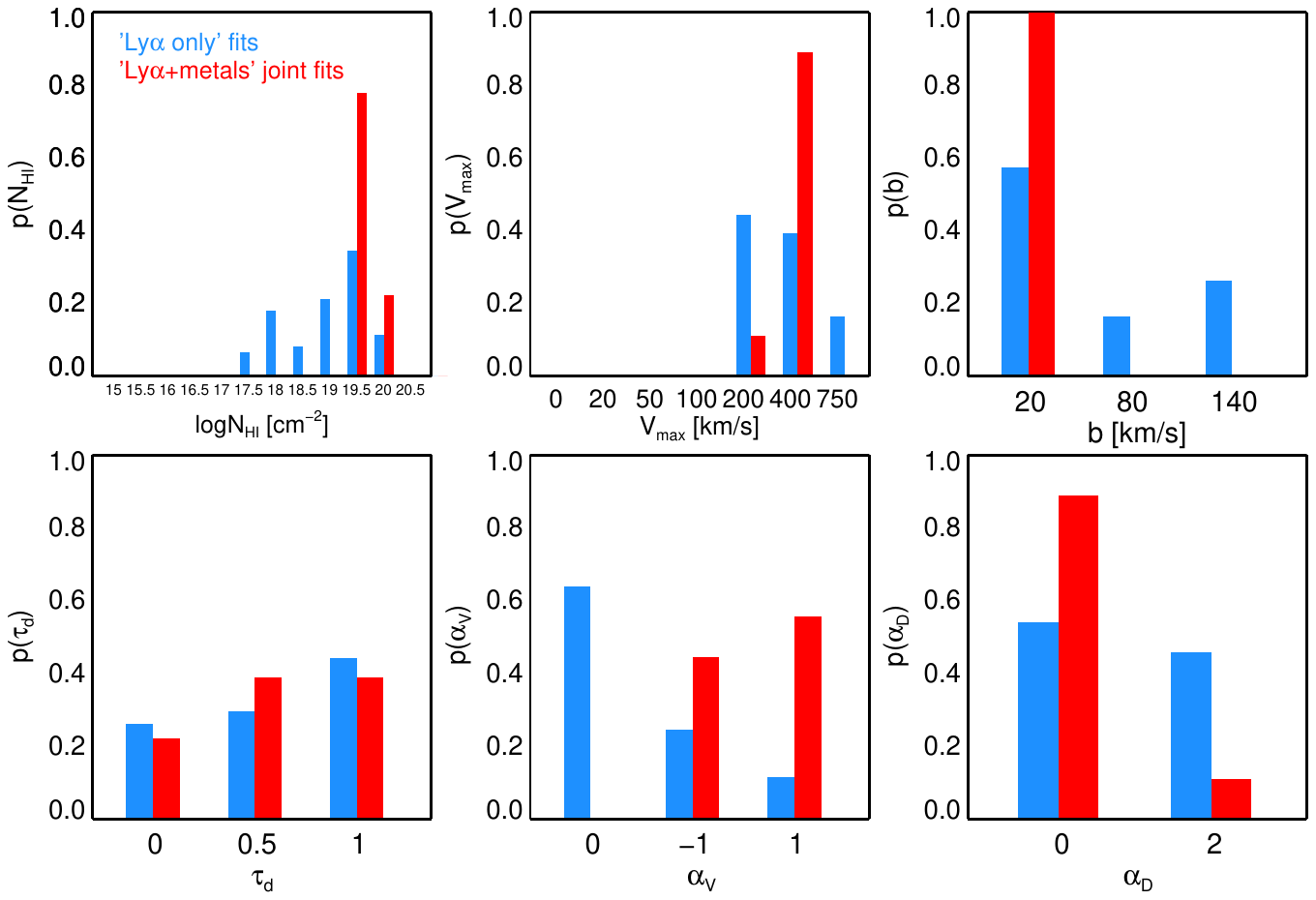}
\vskip-2ex
\caption{Fitting analysis of a MUSE LAE (ID1059) from \citet{Leclercq_2017}. Top panels show the observed \lyat, \siiib{} and \ciil{} lines (grey line with observational uncertainties) with a selection of models from our grid of simulations. Blue curves correspond to best-fit models when applying the fitting procedure to \lya only. The corresponding \siii and \cii lines are given by the wind parameters constrained by the \lya fit. Note that the metal column densities are unconstrained by this procedure, hence we let them take any possible values. Red curves correspond to models that are instead able to fit all lines at once (the thick red curve shows our best fit model). For \ciil, we note that the resonant and fluorescent emissions are blended at MUSE resolution (both in the model and observed spectrum). In each panel, dashed and dotted vertical lines depict the resonant and fluorescent wavelengths respectively. Bottom panels present the posterior distributions of parameter values corresponding to our best fits using \lya only (in blue) and all lines at once (in red).}
\label{fig:jointfit}
\end{figure}

We can further exemplify how our grid of simulations can be used to interpret observations by trying to reproduce jointly observed \lya and metal line profiles. Here, we fit the spectrum of a bright LAE at $z=3.08$ from the MUSE ultra-deep survey \citep[ID1059;][]{Leclercq_2020} which contains \lyat, \siiib{} and \ciil. To do so, we use our full grid of spectra by assuming a Gaussian$+$continuum for the input \lya line and a flat continuum for \siiib{} and \ciil. All simulated spectra have been renormalised to the observed spectra, convolved with the MUSE line spread function (LSF; $\sigma_{\rm LSF} \approx 0.25$ \AA{} at the wavelengths of interest) and degraded to MUSE resolution \citep{Bacon_2017}. When performing the fits, we allow for spectral offsets $\Delta\lambda_{\rm offset} \pm \sigma_{\rm LSF}$ around the line centers in order to account for possible uncertainties on the systemic redshift. Following this procedure, we compute the reduced $\chi^2$ of the three lines for all our simulated spectra and we define a threshold for the goodness-of-fit, $\chi^2_{\rm thresh}$, to extract a sample of models that can well reproduce the observations. We adopt distinct $\chi^2_{\rm thresh}$ values for \lya and metal lines (15 and 2 respectively), owing to their different SNR in the MUSE spectrum of ID1059.

To illustrate the interest of using multiple lines in fitting analyses, we follow a two-step approach to model the observed lines. First, we apply our method only to the \lya line to derive the best-fit parameters, and then we directly predict the corresponding \siiib{} and \ciil{} lines. This is showcased in the top panels of Figure \ref{fig:jointfit} (blue curves).  While the best-fit models well reproduce the observed \lya spectrum, many of them fail to provide a good fit to the metal lines. Next, we now choose to fit jointly \lyat, \siiib{} and \ciil{} and select the sub-group of models that can provide a good match to both \lya and the metal lines ($\chi^2_{Ly\alpha} < \chi^2_{\rm thresh,Ly\alpha}$ and $\chi^2_{\rm metal} < \chi^2_{\rm thresh,metal}$; thin red curves). The best-fit model to all three lines, depicted as a thick red curve, can now nicely reproduce the resonant and fluorescent features of \siiib{} and \ciil{} as well as the typical asymmetric shape of the \lya line.

Looking at the posterior distributions of the `\lya only' fits and `\lyat$+$metals' joint fits (bottom panels of Figure \ref{fig:jointfit}), we find that using multiple lines can help narrow down the allowed range of best-fit parameter values. In our example, this is particularly true for $N_{\rm HI}$, $V_{\rm max}$ and $b$, and to a lesser extent for $\alphav$ and $\alphad$, which become more tightly constrained with the joint fitting procedure. As expected, the dust opacity is found to be the least constraining wind parameter since $\tau_{\rm dust}$ barely affects the shapes of the line profiles in our models (as was shown in Figure \ref{fig:spec_taud}).

\subsection{Role of infilling in shaping metal lines}
\label{subsec:infill}

\begin{figure}
\includegraphics[width=0.59\textwidth,valign=c,trim=0.0cm 3.cm 0.5cm 10.8cm, clip]{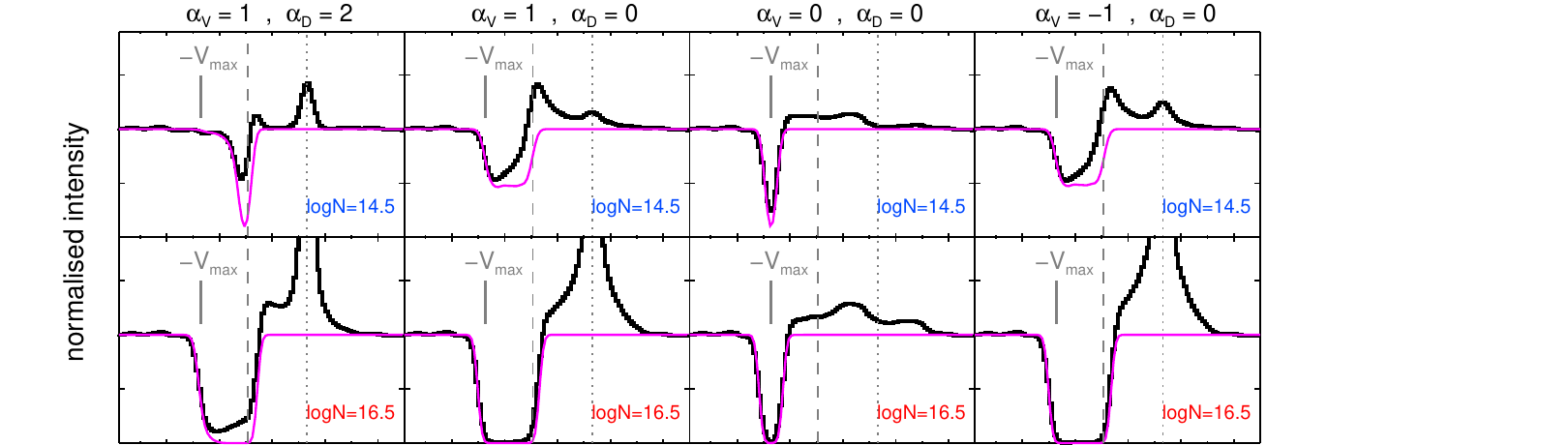}
\includegraphics[width=0.59\textwidth,valign=c,trim=0.0cm 3.cm 0.5cm 11.1cm, clip]{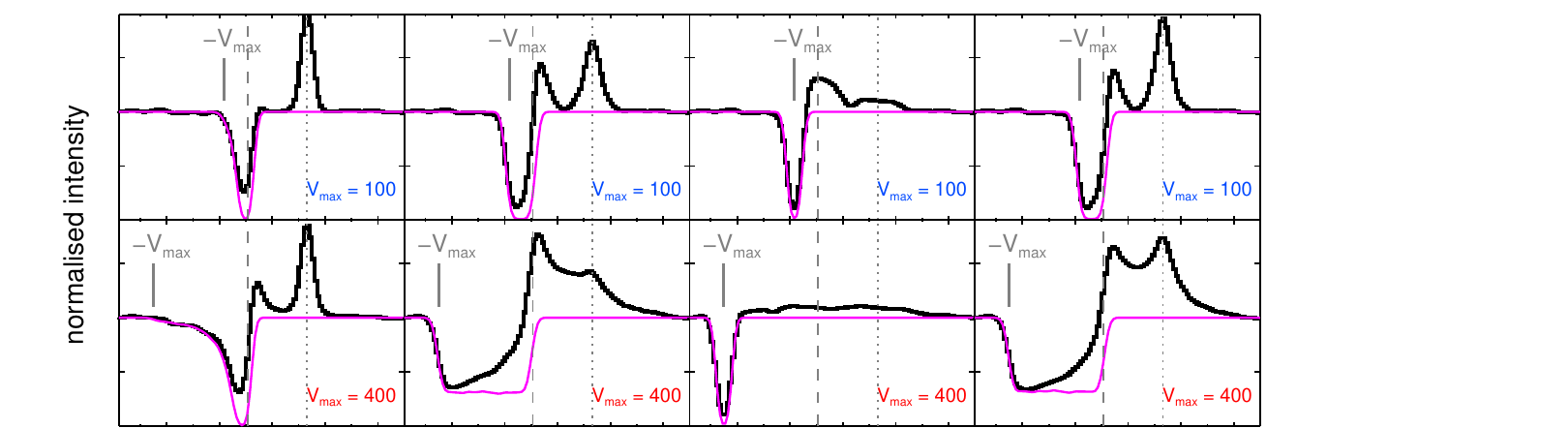}
\includegraphics[width=0.59\textwidth,valign=c,trim=0.0cm 1.3cm 0.5cm 11.1cm, clip]{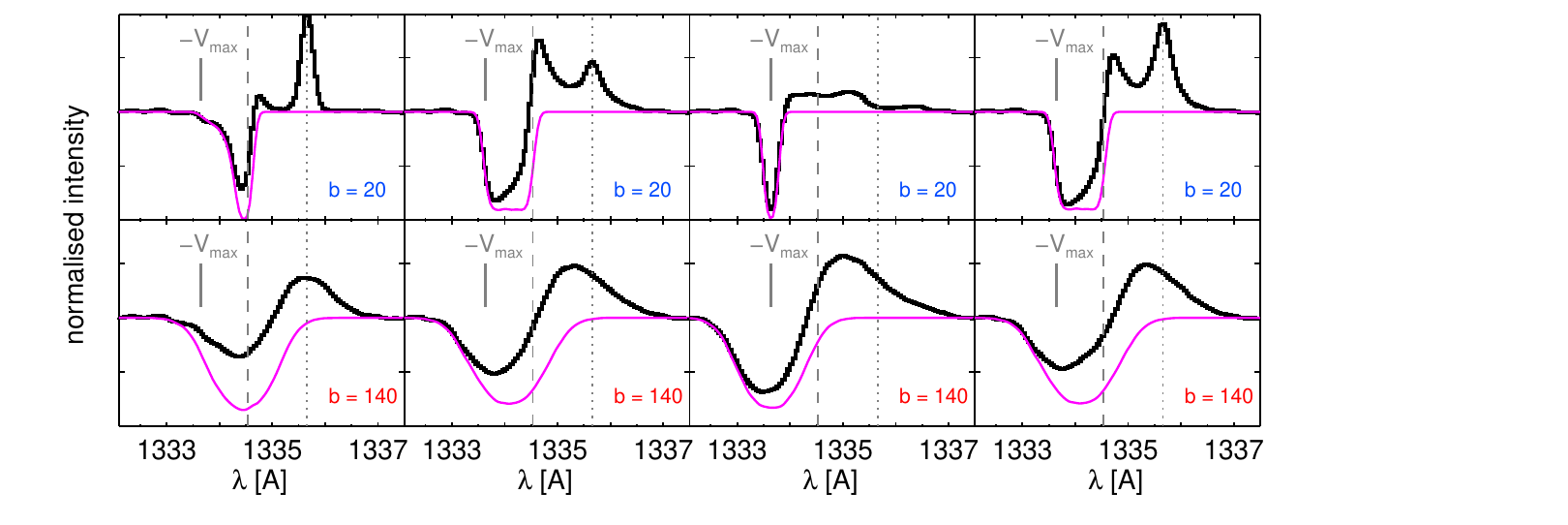}
\caption{Illustration of the impact of the wind parameters on line infilling. Black curves show \ciil{} spectra resulting from the RT simulations whereas magenta lines depict the true absorption profiles (i.e. ignoring re-emission). Columns correspond to various combinations of velocity/density gradients. In all spectra, the continuum is normalised to one. The top, middle, and bottom panels show the spectra for different values of the gas column density, maximum velocity and Doppler parameter respectively. In all cases, the non-varying parameters are fixed to the following fiducial values : log$(N/{\rm cm}^{-2})=15$, $V_{\rm max} = 200$ \kms, $b=20$ \kms, $\tau_{\rm d} = 0$. Dashed (dotted) lines mark the wavelength of the \ciil{} resonance (C\textsc{ii}$^{*}$ $\lambda 1335$ fluorescence). Grey solid lines indicate the wavelength (in velocity units relatively to \ciil) of photons that are seen at line centre by an element of outflowing gas at $V_{\rm max}$.}
\label{fig:infilling}
\end{figure}

When propagated through gas outflows, continuum photons may either escape directly, be reprocessed into a fluorescent channel, destroyed by dust, or scattered around the resonance. In the latter case, photons scattered off the backside of the wind may produce a red emission peak while others may fill-in the absorption on the blue side. Photon infilling can be an important effect that may hamper the interpretation of observed spectra when performing line modelling, Voigt profile fitting or photon conservation analyses \citep{Prochaska2011,Zhu2015,Finley2017}. In our simulations, it can reduce the measured absorption equivalent widths up to 100\% in the most extreme cases.

In order to illustrate its various impacts on LIS lines, we show in Figure \ref{fig:infilling} simulated \ciil{} spectra for various combinations of wind model parameters (black curves) and compare them to the true absorption profiles (magenta curves), i.e. the profiles that would be produced if photons absorbed by the gas were not re-emitted. As shown by the various panels in Figure \ref{fig:infilling}, infilling may shallow the absorption blueward of the resonance, shift the line centroid, or even revert the asymmetry of the blueshifted absorption profile. 

The role of infilling is stronger when the scattering medium is close to the optically thick limit, e.g. for log$(N/{\rm cm}^{-2}) = 14.5$ for \ciit{} (top row of Figure \ref{fig:infilling}). For larger column densities, the line tends to saturate such that all photons are re-absorbed multiple times and will predominantly escape the medium via fluorescent decay (or be absorbed by dust if any). Wind kinematics, characterised by the $V_{\rm max}$ and $b$ parameters, induce Doppler shifts that can favour the frequency redistribution of photons around the resonance. This effect becomes stronger when $V_{\rm max}$ or $b$ are large (bottom panels of Figure \ref{fig:infilling}). As shown by the different columns of Figure \ref{fig:infilling}, infilling can be more or less prominent depending on the radial gas density and velocity profiles. For instance, its effect is usually minimised for constant-velocity models (see third column of Figure \ref{fig:infilling}) because the absorption is mostly centred at $-V_{\rm max}$ whereas photons are re-emitted close to the resonance. In contrast, for accelerated or decelerated winds, photons can be absorbed over a wider frequency range, i.e. between $-V_{\rm max}$ and $0$. In this case, the absorption near the resonance can be more easily refilled and the line centroid may even blueshift towards $V \approx -V_{\rm max}$. 

We note that additional aspects, not accounted for in our simulated spectra, may further complexify the interpretation of line infilling. As emphasised in \citet{Scarlata2015}, absorption lines strictly originate from intervening gas along the line-of-sight whereas scattered photons may instead be re-emitted towards the observer at a transverse distance which is larger than the aperture size of the spectrograph. Unlike our simulated spectra which are computed by collecting all escaping photons, the contribution of infilling in real observed spectra can be significantly reduced due to limited apertures \citep[see also][]{Zhu2015,Xu2022}. Meanwhile, resonant photons may also escape anisotropically (and preferentially along low opacity sightlines) when considering non-spherical or partially-occulted winds. In such cases, scattered radiation may or may not partially fill the underlying absorption line depending on the geometry and the direction of observation \citep{Carr_2018,Mauerhofer2021}.

Overall, our grid of simulated spectra can provide an informative tool to interpret real observations and explore various signatures of line infilling. They may further be used in complement of existing methods to help recover the true underlying absorption profiles in galaxy spectra \citep[e.g.][]{Zhu2015}. 

\section{Discussion}
\label{sec:discussion}

In the previous sections, we have investigated the diversity of signatures onto \lyat, \mgiit, \ciit, \siiit, and \feii line profiles due to the propagation of photons in idealised galactic winds, as predicted by our grid of simulated spectra. Here, we first review the main assumptions of our wind models and possible limitations. Then, we discuss the typical conditions under which the gas is expected to be optically thick to UV lines and to which extent they can be used as probes of ionising radiation leakage.

\subsection{Model limitations and comparisons with the literature}
\label{subsec:caveats}

As described in Section \ref{subsec:models}, our wind model provides an idealised representation in which the central point-source is a proxy for stellar continuum and \lyat/\mgiit{} emissions originating from the ISM, whereas the spherically-symmetric outflow acts as a purely scattering/absorbing medium. In the following, we discuss the main assumptions of this model in the context of recent studies.

According to high-resolution hydrodynamic simulations, resonant lines such as \lya may initially be \textit{pre-processed} in dense star-forming clouds \citep[$T \ll 10^4$ K;][]{Kimm_2019,Kakiichi_2021} and ISM disk \citep{verhamme2012,Smith_2022} before propagating through large-scale outflows.
These small-scale RT effects may broaden the line (which is somehow accounted for in our model by assuming a fixed width for the intrinsic Gaussian emission of $FWHM=150$ \kms), but it may also alter the overall shape of the spectrum and the isotropic escape of photons due to kinematic and geometrical effects in the multiphase ISM.
Additionally, intrinsic continuum emission of real galaxies is likely not arising from a single region but rather comes from the joint contribution of multiple stars distributed across the stellar disk. Likewise, the origin of \lya and \mgiit{} emission lines is not necessarily restricted to a central star-forming region but can also be produced \textit{in-situ} in the CGM by recombining and collisionally-excited gas \citep[e.g.][]{Dijkstra2019,Nelson_2021}. While these mechanisms may be important to explain the most outer regions of observed \lya haloes \citep[e.g.][]{Leclercq_2017,Mitchell_2020b,Niemeyer2022}, simulations by \citet{Byrohl_2021} suggest that rescattered \lya radiation from central star-forming regions is likely the major powering source.

Unlike the spherical and homogeneous gas distribution assumed in our model, galactic winds may instead exhibit anisotropic morphologies with partial covering fractions and/or inhomogeneous structures. For instance, biconical outflows have been extensively studied in the context of \lya and LIS lines \citep{Zheng_2014,Behrens2014,Carr_2018,Burchett2021}. In these models, the radiation escapes mostly 'face-on' and the emergent spectrum strongly varies depending on the viewing-angle and the opening-angle of the cone. Observationally, circumgalactic emission often looks either circular or elongated \citep{Hayes_2014,Wisotzki_2016,Burchett2021,Dutta_2023}. Nevertheless, the exact geometry of the wind is hard to directly infer since we only have access to a projected view of the emitting gas. Still, using stacked \mgii images at $z\approx1$, \citet{Guo2024} recently found evidence for a typical biconical structure around relatively massive galaxies ($M_{\star} \gtrsim 10^{9.5} \msun$), hinting towards the presence of a bipolar outflow perpendicularly to the disk, whereas their low-mass sample exhibits more isotropic emission. In addition, we note that \citet{Zabl_2021} could successfully interpret the absorption and emission features of \mgiit{} halo at $z=0.7$ using a biconical model. In contrast, \citet{Burchett2021} found that spherical geometries better reproduce observations of another \mgiit{} halo at a similar redshift.
Furthermore, our models assume that dust is uniformly mixed with the gas (see Section \ref{subsec:models}) and thus ignore the possibility of a directional variation due to differential dust opacities \citep{Zhang_2023,Cochrane_2024,Gazagnes_2024}. This effect may lead to a large dispersion of the observed properties as a function of the viewing-angle, although it has also been argued that dust attenuation may predominantly occur in star-forming clumps rather than in large-scale outflows, hence mitigating its dependency on inclination \citep{Lorenz_2023}.

Another popular attempt to improve RT models is to include clumpy gas distributions instead of homogeneous fields. As shown in \citet{Gronke_2016}, these models are not always as successful as homogeneous shell models to reproduce the diversity of \lya profiles \citep[but see][]{Erb2023}. Interestingly, it appears that above a critical \textit{covering factor}, clumpy gas distributions display the same behaviour as homogeneous media when applied to e.g. \lya or \mgiit{} \citep{Gronke_2016,Chang2023,Chang24}. Indeed, assuming an optically thick system at fixed integrated column density, the gas can either be made of a limited number of optically thick clouds per sightline embedded in a transparent volume (low-covering factor), or be distributed over numerous, small, optically thin, contiguous clumps (high-covering factor). In the former case, resonant photons propagate via a random walk by scattering off the surface of the clouds and escape along holes. In the latter case, photons escape rather via frequency excursion \citep[i.e. through a series of wing scatterings;][]{adams72}, which mimics the behaviour of radiative transfer in an homogeneous medium and thus yields similar line profiles \citep[see ][for a detailed discussion]{Gronke_2017b}.
 
As illustrated in Figure \ref{fig:allspec}, the gas kinematics are another key ingredient in shaping the line properties. As such, confronting RT model predictions to observations is crucial in the context of galaxy formation and evolution in order to attempt to place constraints on wind dynamics and associated feedback processes. A commonly-used modelling assumes a radial evolution of the wind velocity \citep[e.g.][]{Murray_2005,Heckman_2015}, which may result from the competition between the injection of momentum/energy by radiation pressure or ram pressure forces and gravitational pull from the internal stars and gas. In this picture, the gas can be rapidly accelerated at small radii before reaching a steady-state and/or decelerating at larger distances. Using such models, \citet{Song2020} and \citet{Erb2023} can successfully reproduce \lya spectra along with surface-brightness profiles and LIS absorption lines respectively \citep[see also][]{dijkstra2012b}. In contrast, our grid of simulations explores separately constant-velocity, accelerating, and decelerating outflows in order to highlight the distinguishable signatures associated with these different velocity profiles. 

\subsection{\lya and metal line opacities in the CGM : orders of magnitude}
\label{subsec:line_opacities}

As discussed in Section \ref{sec:lines_set}, \hi and metal ion fractions can be quite low depending on gas temperatures and photoionisation levels. If so, the total gas column densities need to be sufficiently high to make the scattering media optically thick. To assess whether this is the case or not in the surrounding medium of SF galaxies, one should ideally perform specific simulations of the CGM. While such detailed analysis is well beyond the scope of this paper, we can try to estimate the typical \lya and metal ion opacities using simple arguments based on virial scaling relations and the \cloudy predictions presented in Section \ref{sec:lines_set}.

\begin{figure}
\includegraphics[width=0.24\textwidth]{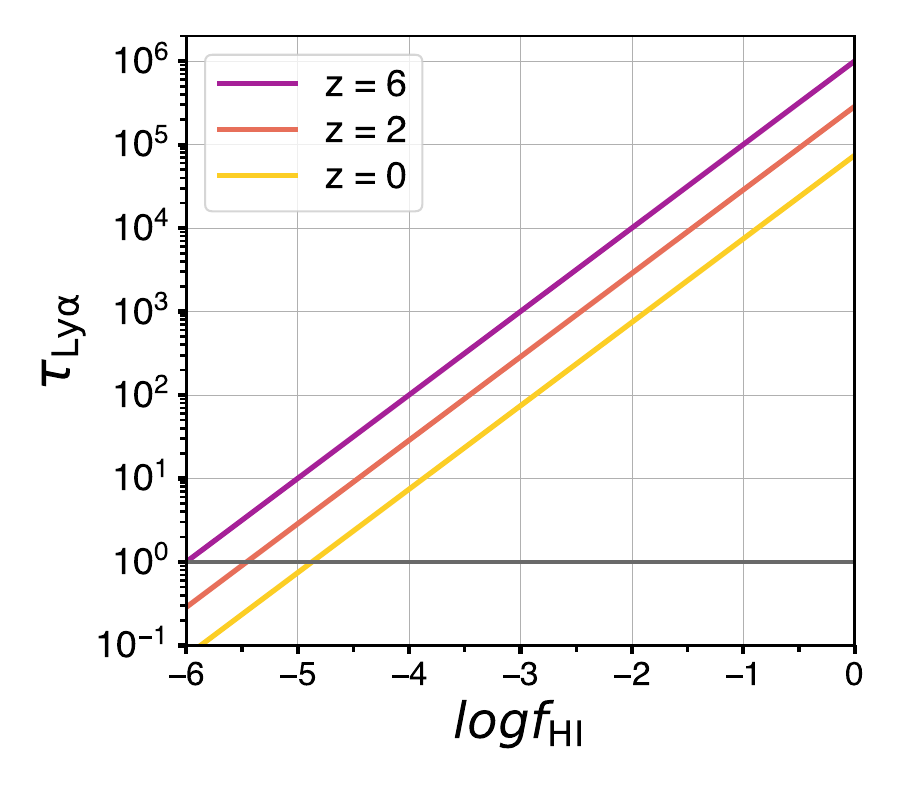}
\includegraphics[width=0.24\textwidth]{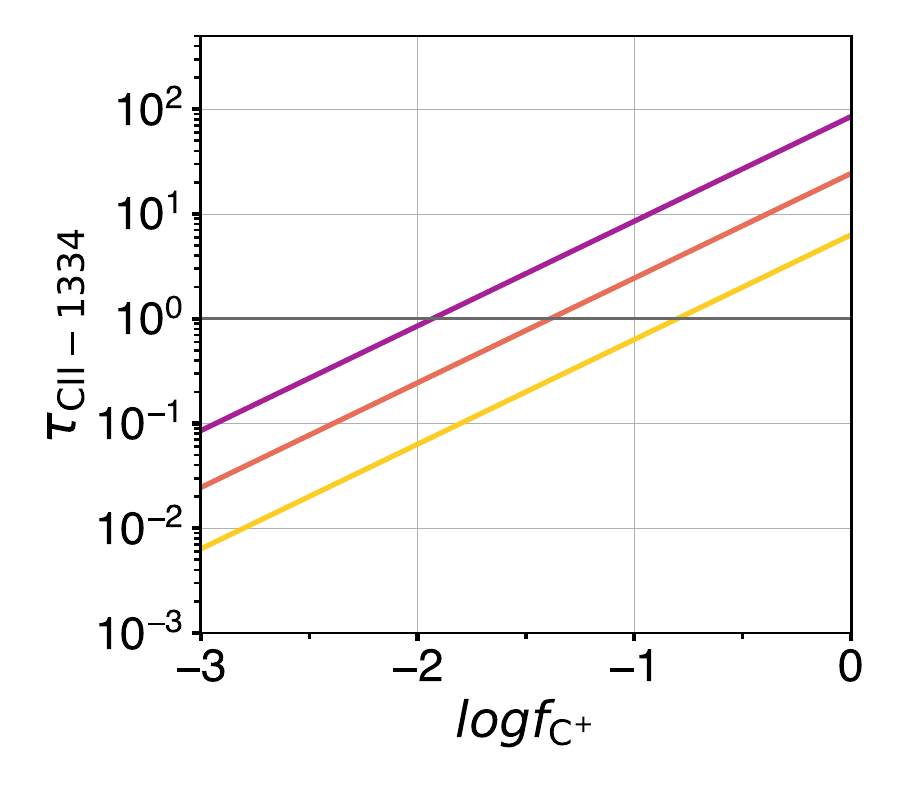}\\
\includegraphics[width=0.24\textwidth]{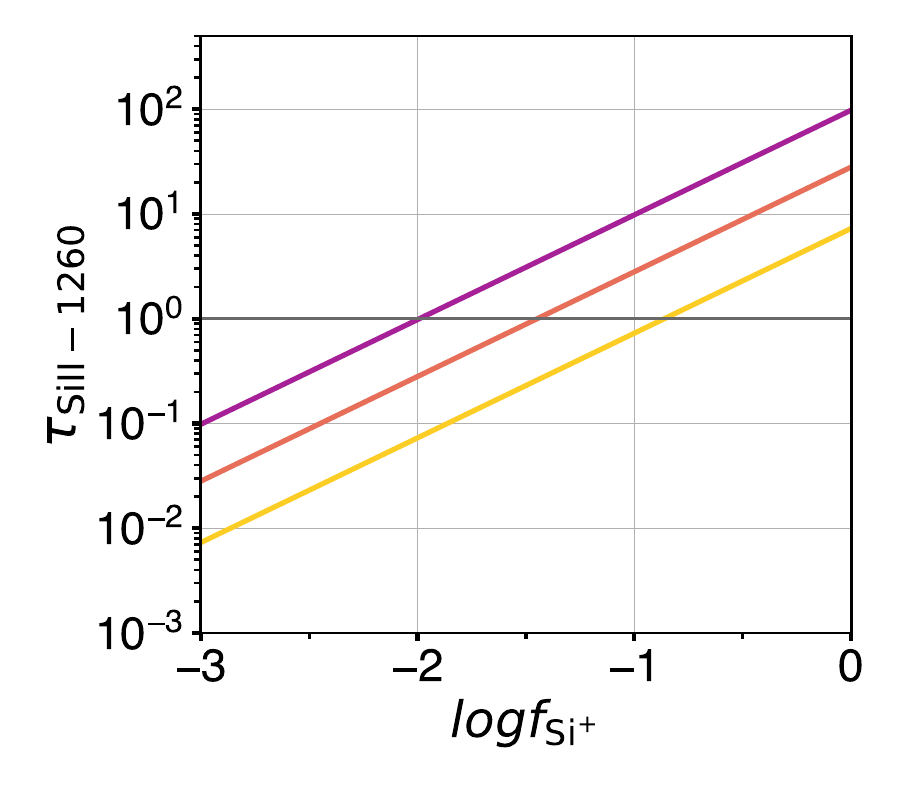}
\includegraphics[width=0.24\textwidth]{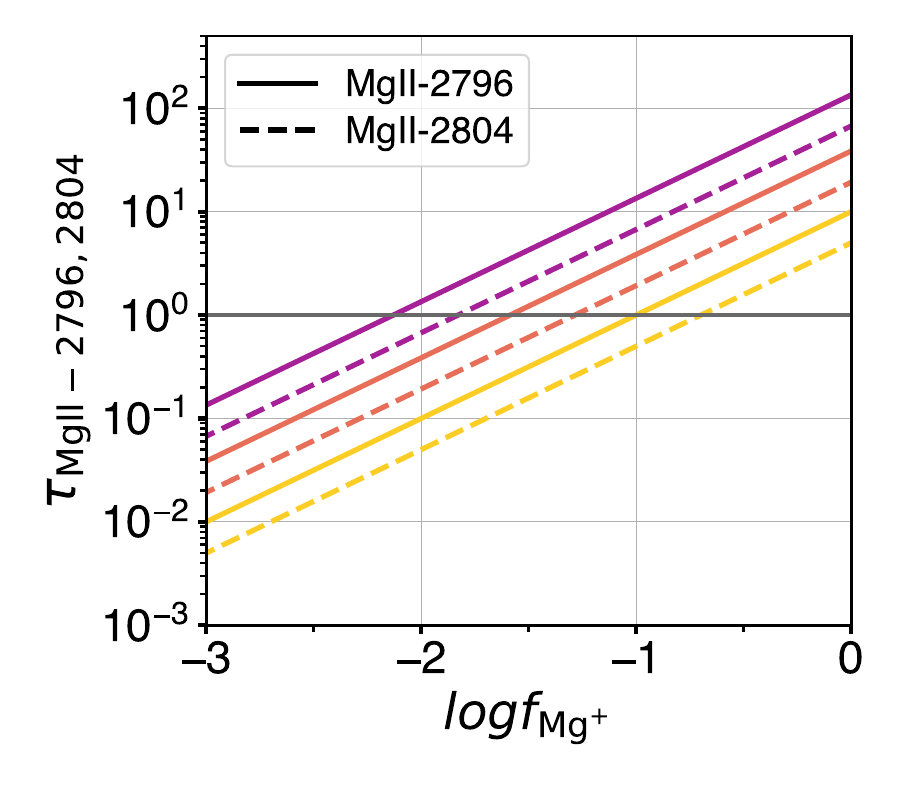}\\
\includegraphics[width=0.24\textwidth]{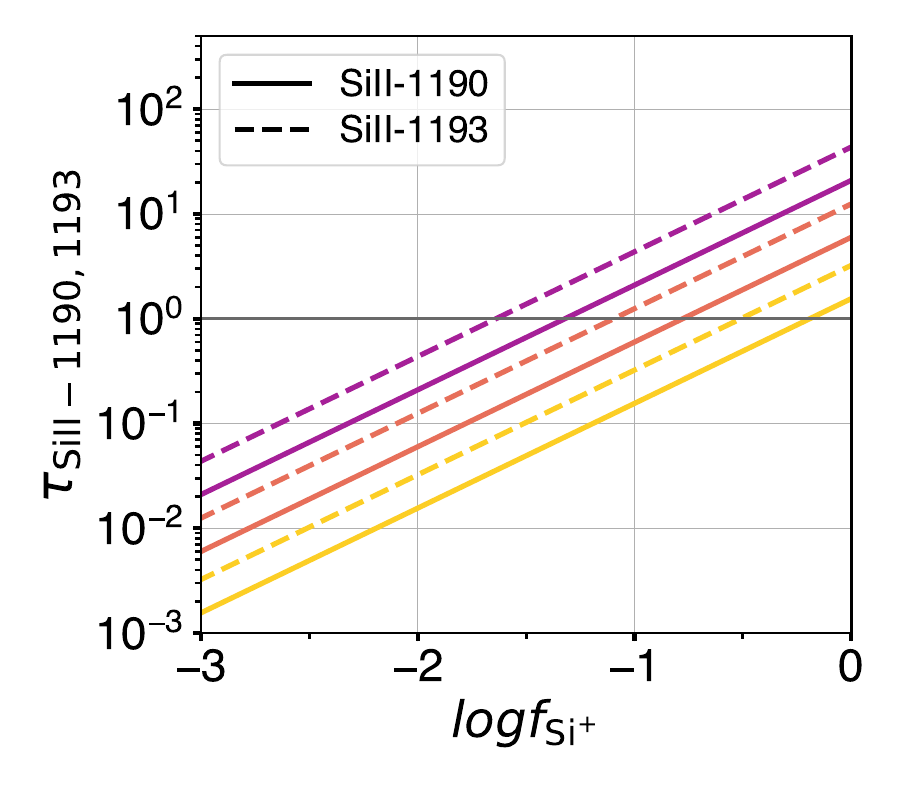}
\includegraphics[width=0.24\textwidth]{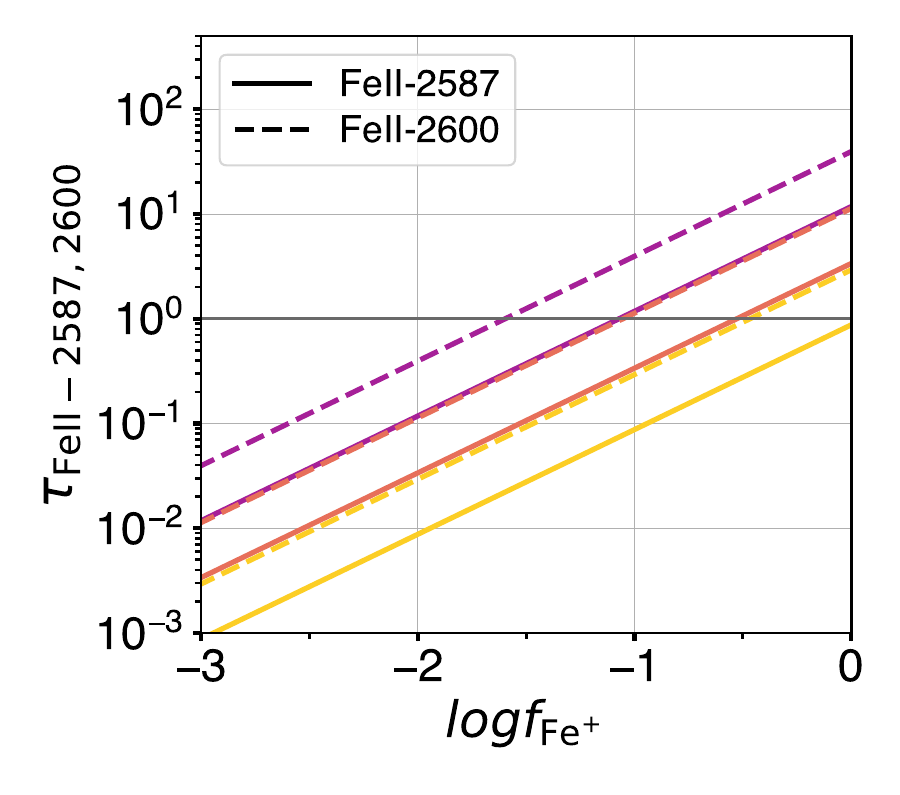}
\caption{\lya and metal line opacities as a function of \hi neutral fraction and singly-ionised fraction respectively. Opacities are computed from Eq. \ref{app_eq6} in Appendix \ref{appendix2} assuming $f_{\rm CGM}=0.25$, primordial hydrogen abundance and solar neighbourhood ratios (see Table \ref{table_1}). The different coloured curves correspond to $z=0, 2$ and 6. Dashed curves show the longer wavelength resonant transition in case of doublets/multiplets.}
\label{fig:lis_opacities}
\end{figure}

To do so, we resort to a toy model based on scaling arguments which uses the baryonic content of DM haloes as a proxy for the amount of gas in the CGM. As described in Appendix \ref{appendix2}, we assume that the CGM i) extends from the halo centre to the virial radius, ii) contains a fraction $f_{\rm CGM}$ of the total baryonic content of the halo, and iii) has a uniform density distribution. 
Using virial relations, we can estimate the gas properties (i.e. the gas scattering cross-section and column density) from the average halo density, radius, and temperature to compute the gas opacity,  $\tau_{\rm g}$ (see Eq. \ref{app_eq6}), at line centre (i.e. ignoring the effect of wind kinematics). In this framework, $\tau_{\rm g}$ is found to be independent of the halo properties and only varies with redshift (see Appendix \ref{appendix2}).
Assuming a primordial hydrogen abundance and solar metallicity, the \lya and LIS line opacities are given by $\tau_{Ly\alpha} = 0.76 f_{\rm HI} \tau_{\rm g}$ and $\tau_{X} \approx 0.76 f_{\rm X^{+}} (X/H)_{\odot} \tau_{\rm g}$ respectively. We show the results for each line in Figure \ref{fig:lis_opacities}.   

In highly neutral gas ($f_{\rm HI} \approx 0.1-1$), \lya opacities can reach values up to $10^5-10^6$, corresponding to extremely thick media. In this regime, the effect of resonant scattering is predominant as often inferred from simulated \lya line profiles and spatially-extended emission \citep[e.g.][]{verh08,barnes2011a, zheng2010a,Byrohl_2021,Camps_2021,Blaizot2023}. In addition, the medium is likely to remain opaque to \lya even in almost fully ionised gas, i.e. $\tau_{\rm Ly\alpha} \approx 1$ when $f_{\rm HI} \lesssim 10^{-5}$. We showed in Figure \ref {fig:ionfrac} that the \hi fraction is always above $10^{-5}$ for gas temperatures of $T \approx 10^{4}-10^{5}$ K. Consequently, the scattering media around SF galaxies are expected to always be optically thick to \lya radiation, even in warm gas phases (i.e. $T \approx10^{5}$ K), at any redshift considered here.

Similarly to \hit, the metal ion opacities increase with ionised fractions and redshift. Nevertheless, they are restricted to much lower values and span a range of $10^{-3} \lesssim \tau_{X} \lesssim 10^2$ (assuming solar abundances) which is broadly similar to the typical values inferred from observations and models/simulations \citep{Martin_2009,Prochaska2011,Nelson_2021}. Based on Figure \ref{fig:ionfrac}, metal ion fractions can vary significantly depending on gas temperatures and incident radiation intensities. In the range $T\approx 10^{4}-10^{5}$ K, $f_{\rm X^{+}}$ can drop from 1 to $\approx 10^{-4}$. Meanwhile, Figure \ref{fig:lis_opacities} tells us that the optically thick limit is only reached for $f_{\rm X^{+}} \gtrsim 0.01$ at $z=6$ and $f_{\rm X^{+}} \gtrsim 0.1$ at $z=0$. Therefore, most metal lines considered in this study are expected to be in the optically thick regime only when the temperature is close to $T=10^4$ K, i.e. where $f_{\rm X^{+}}$ is maximal. Still, we note that \cplus{} fractions are generally above 10\% at $T\approx 10^{4}-10^{5}$ K (see Figure \ref {fig:ionfrac}) so most media are always optically thick to \cii lines in this temperature range, such that \ciil{} may be tracing gas over the same temperature range as \lyat. In contrast, \feiid{} lines are the least likely to be affected by resonant scattering in such media since \feplus{} fractions quickly drop at $T > 10^4$ K while their optically thick limit requires $f_{\rm X^{+}}$ of the order of unity at $z=0$ and $\approx 0.1$ at $z=6$.

Keeping in mind the toy-model nature of our CGM opacity estimates, our analysis nevertheless suggests that \lya and \ciit{} are good tracers of cool/warm gas over the full temperature range $T \approx 10^4-10^5$ K. In contrast, the use of the other LIS lines may be restricted to probe cooler phases, i.e. with temperatures around $T \approx 10^4$ K.

\subsection{LIS lines as tracers of LyC leakage}
\label{subsec:lyc_leakage}

Being mostly sensitive to gas phases containing neutral hydrogen (Figure \ref{fig:ionfrac}), UV resonant lines are promising probes of ionising radiation leakage from high-redshift galaxies or low-redshift analogues \citep{Steidel_2018,Chisholm2020,Mauerhofer2021,Saldana_Lopez_2022,Gazagnes_2024,Leclercq_2024}. With an optically thin limit of $N_{\rm HI} \approx 10^{17} \: {\rm cm}^{-2}$ \citep{osterbrock2006}, LyC escape occurs along low \hi column density channels. Therefore, looking for resonant scattering signatures in lines that probe \hi gas phases (e.g. blueward absorption and/or P-Cygni profiles, fluorescent emission) may help distinguish between leaking and non-leaking sources in the density-bounded regime \citep[][]{Zackrisson2013,Jaskot2014}.

Expressing the \lya opacity at line center as $\tau_{\rm Ly\alpha}=(\sigma_{\rm Ly\alpha}/{10^{\scaleto{-13}{4.5pt}} {\rm \: cm}^2}) (N_{\rm HI}/10^{13} \: {\rm cm}^{-2})$ (see Section \ref{sec:lines_set} and Table \ref{table_1}), the medium is expected to be moderately opaque to \lya radiation in LyC leakers (i.e. $\tau_{\rm Ly\alpha} \lesssim 10^3-10^4$ at $N_{\rm HI} \lesssim 10^{17}-10^{18} \: {\rm cm}^{-2}$). In this regime, the escape of ionising photons has been shown to correlate well with various \lya line properties such as peak separation and equivalent width \citep[][]{Verhamme2017,Steidel_2018,Choustikov_2024}. While usually less prominent than \lya in galaxy spectra, LIS lines also hold great potential for the search of ionising sources, especially during the epoch of reionisation where the observability of the \lya line is strongly hampered by IGM attenuation \citep[][]{Garel2021}. To probe faithfully LyC leakage, an ideal line should exhibit reduced (or null) signatures of resonant scattering when propagated in gas with $N_{\rm HI} \lesssim 10^{17} \: {\rm cm}^{-2}$. In other words, its opacity should scale exactly as the LyC opacity with \hi column density. Here, we explore the cases of \mgiit, \ciit, \siiit, and \feiit{} to assess to which extent they can be seen as reliable observational probes of LyC escape.

To first order, the line opacity of a metal ion can be written as a function of the ratio of the singly-ionised metal fraction and \hi neutral fraction ($f_{\rm X^+}/f_{\rm HI}$), the abundance ratio relative to hydrogen $(X/H)$, and the dust depletion $\delta_{\rm X}$:
\begin{equation}
\label{eq:taux_nhi}
\begin{aligned}
\tau_{\scaleto{\rm X}{3.9pt}} \: ={} & \left(\cfrac{\sigma_{\rm X,0}}{10^{-13} \: {\rm cm}^2}\right) \left(\cfrac{b}{20 \: \kms }\right)^{-1} \left(\cfrac{f_{\rm X^+}}{f_{\rm HI}}\right) \\ & \qquad \qquad \qquad \qquad \left(\cfrac{X/H}{10^{-4}}\right) (1-\delta_{\rm X}) \left(\cfrac{N_{\rm HI}}{10^{17} \: {\rm cm}^{-2}}\right)
\end{aligned}
\end{equation}
where $\sigma_{\rm X,0}$ are the line centre cross-sections for each transition (see Table \ref{table_1}).  According to solar neighbourhood ratios, $(X/H)$ is of the order of $10^{-4}-10^{-5}$ for Mg, Si, Fe, C (Table \ref{table_1}) and can be estimated with relatively good accuracy from observations. A main unknown in Eq. \ref{eq:taux_nhi}, often poorly constrained in observations, is the dust depletion which can substantially vary from one galaxy to another. Based on MW, LMC and SMC observations \citep{Roman_Duval_2022}, $\delta_{\rm X}$ can reach values $\gtrsim 80-90\%$ for iron, $\approx 20-70\%$ for silicon and magnesium, and $\approx 20-30\%$ for carbon (see Table \ref{table_1} for fiducial values $\delta_{\rm X,0}$). Another critical parameter in Eq. \ref{eq:taux_nhi} is $(f_{\rm X^+}/f_{\rm HI})$, the ratio of the singly-ionised metal fraction and \hi neutral fraction. In the literature, this ratio is often taken to be of the order of unity when estimating metal line opacities as a function of $N_{\rm HI}$ \citep[e.g.][]{Chisholm2020}. This choice implicitly assumes a very cold phase, typical of ISM gas (i.e. $T \ll 10^4$ K), and/or neglects the contribution of higher-order ionisations of hydrogen and metals. However, $(f_{\rm X^+}/f_{\rm HI})$ is highly unconstrained and model-dependent as demonstrated in Figure \ref{fig:ionfrac} where we had shown that it can basically take any value in the range $\approx 1-100$ at $T \approx 10^4$ K.

\begin{figure}
\hskip0.5ex
\includegraphics[width=0.51\textwidth,valign=c,trim=3.9cm 1.cm 2.9cm 0.8cm, clip]{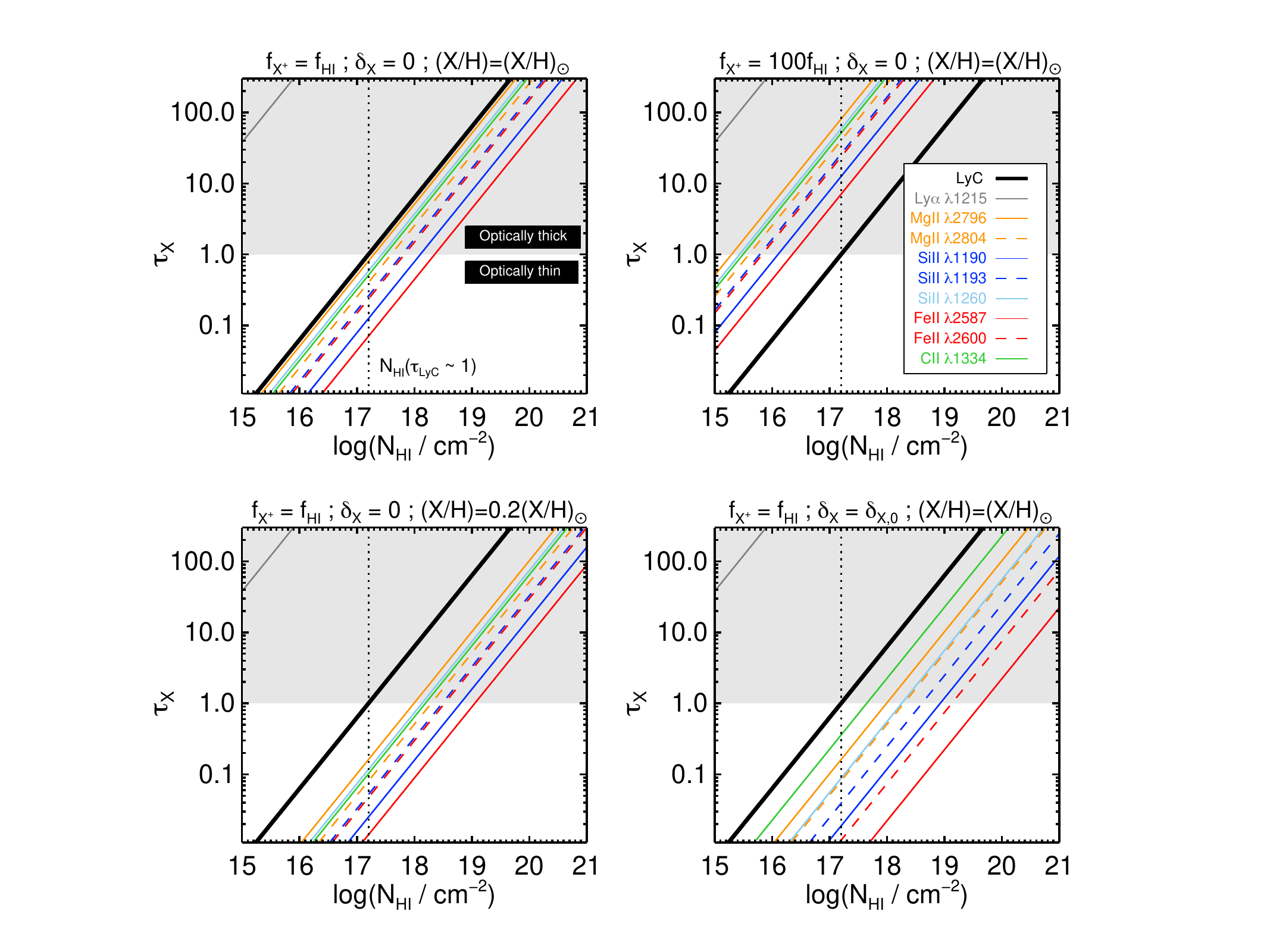}
\caption{Comparison of the LyC opacity (thick black line) with the line centre opacities of each transition (thin coloured lines) considered in this study as a function of \hi column density. The gas opacity of each line is estimated according to Eq. \ref{eq:taux_nhi} and assumes solar abundances $(X/H)=(X/H)_{\odot}$. Each panel corresponds to different sets of dust depletion $\delta_{\rm X}$, metallicity $(X/H)$, and relative singly-ionised metal fraction to \hi neutral fraction $(f_{\rm X^{+}}/f_{\rm HI})$ values. $\delta_{\rm X,0}$ is the dust depletion based on MW measurements (see text and Table \ref{table_1}). The \taux{} $ = 1$ horizontal line marks the transition between optically thin and thick regimes. The dotted vertical line depicts the Lyman limit (i.e. $N_{\rm HI}=1.6\times10^{17}$ cm$^{-2}$) below which ionising photons can efficiently escape the medium.}
\label{fig:taux_nhi}
\end{figure}

Figure \ref{fig:taux_nhi} shows the impact of the parameters of Eq. \ref{eq:taux_nhi} on the relations between metal line and LyC opacities with $N_{\rm HI}$. In the case where $f_{\rm X^+}=f_{\rm HI}$, $\delta_{\rm X}=0$ and $(X/H)=(X/H)_{\odot}$ (top left panel), the optical thin limit of the metal lines is reached at $(N_{\rm HI}/$cm$^{-2}) \approx 10^{17}-10^{18}$, which is broadly consistent with that of LyC. However, if other plausible parameter values are assumed (other panels of Figure \ref{fig:taux_nhi}), metal line and LyC opacities scale very differently with $N_{\rm HI}$. According to our results, the transition between optically thin and thick LIS lines is scattered over nearly five orders of magnitudes of \hi column densities, from $N_{\rm HI} \approx 10^{15}$ to $\approx 10^{20}$ cm$^{-2}$. In other words, metal line opacities can strongly underestimate or overestimate the LyC optical depth, such that metal lines may no longer be used to accurately infer LyC leakage. Therefore, optically thin metal lines may or may not trace LyC-leaking systems depending on the actual ionised fractions, gas metallicity or dust depletion. 

The complex connection between LyC leakage and LIS line properties has also been highlighted by recent hydrodynamical simulations based on more realistic setups/assumptions than ours \citep[e.g. accounting for anisotropic gas geometries, dust attenuation, etc;][]{Mauerhofer2021,Katz2022,Gazagnes2023}. Altogether, this suggests that inferring reliable and quantitative information on the LyC leakage from individual sources based on LIS lines is probably very challenging \citep[e.g.][]{Chisholm2020}. Still, metal gas opacities are expected to roughly scale like \hi opacities in most cases, albeit with significant scatter (Figure \ref{fig:taux_nhi}), such that these limitations may be partly overcome using galaxy samples with robust constraints on gas metallicity, ionisation fractions, and/or dust depletion so as to mitigate the expected dispersion in the $N_{\rm HI}-\tau_{\scaleto{\rm X}{3.9pt}}$ relation.

\section{Summary and conclusions}
\label{sec:conclusion}

In this study, we have presented a large grid of $\approx 20,000$ Monte Carlo radiative transfer simulations in $5,760$ wind configurations using a custom version of the \rascas code. 

With this dataset we can predict the spectral properties of \hi \lya and five ultraviolet metal lines associated with \mgplus, \siplus, \cplus, and \feplus{} after radiative transfer in idealised gas outflows with or without dust. Assuming a central source emission, we have considered the continuum pumping scenario in which continuum radiation may be reprocessed by resonant transitions (some being associated with fluorescent decay channels), as well as intrinsic emission lines in the case of \lya and \mgiit. All simulated spectra have been compiled into a public grid which is available through a web portal. In Appendix \ref{appendix1}, we have described the structure of the database and how to use/download the grid of spectra.  

We have performed a broad exploration of our grid to assess the role of our wind parameters, namely the gas column density, wind velocity, Doppler broadening parameter, dust content, as well as density and velocity gradients, in shaping the emergent lines. Our results can be summarised as follows:
\begin{itemize}
   \item[$\bullet$] The spectra exhibit a wide variety of profiles and all wind parameters seem to alter resonant features both in absorption and emission, as well as the fluorescent lines associated with \ciit, \feiit, and \siiit.\\[-5pt]
   \item[$\bullet$] To illustrate the added-value of multiple-line analyses to infer galactic wind properties, we have performed a joint modelling of \lyat, \ciit, and \siii lines of a star-forming galaxy observed at high redshift with MUSE. We argue that this kind of approach can place refined constraints on the underlying wind parameters which may help break some degeneracies that exist among various models.\\[-5pt]
    \item[$\bullet$] In our models, line infilling acts differently depending on wind kinematics, radial velocity and density profiles, and its effect is likely maximal for moderately dense media.
  \end{itemize}
 
We have further discussed the merits of using \lya and metal lines so as to probe the properties of gas outflows and ionising radiation leakage: 
\begin{itemize}
   \item[$\bullet$] Using \cloudy simulations, we have shown that neutral hydrogen and singly-ionised metal ions should always co-exist in typical gas outflows with temperatures of $T \approx 10^4-10^5$ K, although the abundance of these species is maximal in the cooler phase, i.e. around $T \approx 10^4$ K. Based on virial scaling relations, we argue that \lya can probe any gas phase between $T=10^4$ and $10^5$ K, even if the medium is highly ionised (e.g. $f_{\rm HI} \approx 10^{-5}$). While \cii is found to trace nearly the same phase as \lyat, only gas at $T \approx 1-2 \times 10^4$ K may be able to sustain singly-ionised metal fractions that are high enough in order to be optically thick to \feiit, \mgiit, and \siii lines.\\[-5pt] 
   \item[$\bullet$] Our modelling suggests that the absence (presence) of radiative transfer signatures on LIS lines may help probe the transition between LyC leaking (non-leaking) galaxies, i.e. log$(N_{\rm HI}/$cm$^{-2}) \approx 17-17.5$. Nevertheless, we caution that this connection is highly dependent on the assumed \mgplus, \siplus, \cplus, and \feplus{} ionisation fractions, as well as dust depletion and gas metallicity. As such, our modelling predicts that optically thin LIS lines can trace \hi column densities scattered over nearly five orders of magnitude around the LyC limit ($N_{\rm HI} \approx 10^{17}$ cm$^{-2}$) which may restrict the use of metal lines as probes of ionising leakage.
      \end{itemize}

The grid of simulated spectra presented in this paper is intended to analyse self-consistently various resonant lines that are often detected in the UV spectra of star-forming galaxies. As such, it provides a valuable resource that can help interpret observational data and improve our understanding of gas outflows as well as the connection between resonant line RT signatures and the escape of ionising radiation. Looking ahead, this study paves the way for future developments such as the inclusion of additional wind models/parameters, a better sampling of the parameter space and/or new UV lines.

\section*{Acknowledgements}
TG and AV acknowledge support from the SNF grant PP00P2$\_$211023. VM acknowledges support from the NWO grant 0.16.VIDI.189.162 (ODIN). We gratefully acknowledge support from the PSMN (P\^ole Scientifique de Mod\'elisation Num\'erique) of the ENS de Lyon, the Common Computing Facility (CCF) of the LABEX Lyon Institute of Origins (ANR-10-LABX-0066), the CC-IN2P3 Computing Centre (Lyon/Villeurbanne - France, and the LESTA cluster hosted at the Department of Astronomy of Geneva University for the computing resources.

\bibliographystyle{aa}
\bibliography{biblio}


\appendix

\section{Online interface}
\label{appendix1}

Our online database containing integrated spectra for all possible combinations of wind model parameters is accessible online at \url{https://rascas.univ-lyon1.fr/app/idealised_models_grid/}. Each simulated spectrum has been computed either from a flat continuum or a continuum$+$Gaussian emission (for \lya and \mgiid{} only) and with/without LSF convolution. From our online interface ({\tt "Access the grid"} button), users can easily select models according to the following parameters:

\begin{itemize}
    \item[\ding{114}] \textbf{Wind model} : choose models based on the wind parameter values:
\begin{lstlisting}[language=html]
<@\textcolor{teal}{\# Gas column density logN [atoms/cm$^2$] \vspace{0.2cm}}@>
<@\textcolor{black}{$\tt logN = [13.5 , 14, 14.5, 15 , 15.5, $}@>
                         <@\textcolor{black}{$\tt 16, 16.5, 17]$}@> <@\textcolor{black}{\small \!\!\!\! for metals \vspace{0.1cm}}@>
<@\textcolor{black}{$\tt logN = [15 , 15.5, 16, 16.5 , 17, 17.5, 18, 18.5, $}@>
                         <@\textcolor{black}{$\tt 19, 19.5, 20, 20.5, 21]$}@> <@\textcolor{black}{\small \!\!\!\! for HI \vspace{0.01cm}}@>
<@\textcolor{teal}{ \# Density profile $\alpha_{\rm D}$ \vspace{0.2cm}}@>
<@\textcolor{black}{$\tt alphaD\!=\![0 , 2]$}@> <@\textcolor{black}{\small \! \# \!\!\! uniform / isothermal \vspace{0.3cm}}@>
<@\textcolor{teal}{\# Maximum wind velocity $V_{\rm max}$ [km/s] \vspace{0.2cm}}@>
<@\textcolor{black}{$\tt Vmax\!=\![0 , 20, 50, 100, 200, 400, 750]$ \vspace{0.3cm}}@>
<@\textcolor{teal}{\# Velocity profile $\alpha_{\rm V}$ \vspace{0.2cm}}@>
<@\textcolor{black}{$\tt alphaV\!=\![0 , 1, -1 ]$}@><@\textcolor{black}{\small \: \# \!\!\! constant / accelerating / decelerating \vspace{0.2cm}}@>
<@\textcolor{teal}{\# Doppler parameter b [km/s] \vspace{0.2cm}}@>
 <@\textcolor{black}{$\tt b\!=\![20, 80, 140]$\vspace{0.3cm}}@>
<@\textcolor{teal}{\# Dust opacity $\tau_{\rm d}$ \vspace{0.2cm}}@>
<@\textcolor{black}{$\tt taud\!=\![0 , 0.5 , 1]$\vspace{0.3cm}}@>
\end{lstlisting}
     \item[\ding{114}] \textbf{Lines} : choose one or several lines 
     \begin{lstlisting}[language=html]
<@\textcolor{purple}{\# Ly$\alpha$ $\lambda1216$ \vspace{0.2cm}}@>
<@\textcolor{purple}{\# CII $\lambda1334$ \vspace{0.2cm}}@>
<@\textcolor{purple}{\# FeII $\lambda\lambda2586-2600$ \vspace{0.2cm}}@>
<@\textcolor{purple}{\# MgII $\lambda\lambda2796-2803$ \vspace{0.2cm}}@>
<@\textcolor{purple}{\# SiII $\lambda\lambda1190-1193$ \vspace{0.2cm}}@>
<@\textcolor{purple}{\# SiII $\lambda1260$ \vspace{0.2cm}}@>
 \end{lstlisting}
    \item[\ding{114}] \textbf{Input spectrum} : choose the intrinsic spectrum of the source 
         \begin{lstlisting}[language=html]
  <@\textcolor{violet}{[`Flat spectrum',`Gaussian line']}@><@\textcolor{black}{\small \! \# For `Gaussian line', the intrinsic spectrum is a flat continuum plus a Gaussian line centered \:
on the wavelength of the resonant line of interest. The intrinsic line has an equivalent width of 100 \AA{} for Ly$\alpha$ and 6/3 \AA{} for the 2796/2803 MgII doublet lines. The FWHM of each line is set to 150 km/s.}@>
  <@\textcolor{black}{\vspace{-0.1cm}}@>
   \end{lstlisting}
    \item[\ding{114}] \textbf{LSF convolution} : choose spectra with and/or without convolution with a Line Spread Function 
         \begin{lstlisting}[language=html]
  <@\textcolor{orange}{[`Without',`With']}@><@\textcolor{black}{\small \! \# if `With', Gaussian smoothing (0.2 \AA) is applied to the output spectrum.}
  \end{lstlisting}
\end{itemize}

The spectrum files are written in ascii format with two entries: rest-frame wavelengths (\AA) and fluxes (normalised to the continuum). All spectra are binned at very high resolution ($\Delta\lambda = 0.05$ \AA) and can be easily resampled to coarser values. The selection can be downloaded as tarball file (click the {\tt "Download Selection"} button). Alternatively, one may choose to download the full library of integrated spectra as a zipped tarball file ($\approx 250$ MB; {\tt "Download full archive"} button). The unzipped data is about 1.4 GB. 

When downloading the full library or only a subsample of chosen models, the data is organised in main directories corresponding to the wind models (e.g. {\tt \small b20.0\_alphaV1\_alphaD2\_Vmax200.0\_logN19.0\_taud0.5}/), each containing one or several subdirectories corresponding to the selected line(s) (e.g. {\tt \small Lya/}). These subdirectories contain one or several spectrum files depending on the choices of input spectrum and LSF convolution (e.g. {\tt \small spectrum\_FWHM150\_EW100.dat} and {\tt \small spectrum\_LSF0.2A.dat}).
The structure can be summarised as follows: : 
     \begin{lstlisting}[language=html]
<@\textcolor{teal}{\small ---b20.0\_alphaV1\_alphaD2\_Vmax200.0\_logN19.0\_taud0.5/ \vspace{0.02cm}}@>
  <@\textcolor{purple}{\small ---Lya/ \vspace{0.02cm}}@>
    <@\textcolor{gray}{\tiny ---spectrum.dat\hspace{-0.5cm}}@><@\textcolor{black}{\tiny \! \! \! \! \! \! \! \# \hspace{-0.2cm} Flat input \& no LSF \vspace{0.cm}}@>
    <@\textcolor{gray}{\tiny ---spectrum\_FWHM150\_EW100.dat\hspace{-0.5cm}}@><@\textcolor{black}{\tiny \! \! \! \! \! \! \! \# \hspace{-0.2cm} Gaussian input \& no LSF \vspace{0.cm}}@>
    <@\textcolor{gray}{\tiny ---spectrum\_LSF0.2A.dat \vspace{0.02cm}}@><@\textcolor{black}{\tiny \! \# \!\!\! Flat input \& LSF \vspace{0.cm}}@>
    <@\textcolor{gray}{\tiny ---spectrum\_FWHM150\_EW100\_LSF0.2A.dat\hspace{-0.5cm}}@><@\textcolor{black}{\tiny \! \! \! \! \! \! \! \# \hspace{-0.2cm} Gaussian input \&  LSF \vspace{0.3cm}}@>
 \end{lstlisting}
 
Additional simulated data, such as spatially-resolved spectra, surface brightness profiles, escape fractions, properties of individual photons (e.g. frequency, number of scatterings, etc) or integrated spectra with supplementary intrinsic FWHM or EW values, are not yet available through the online interface but may also be provided upon reasonable request at \url{thibault.garel@unige.ch}. 

\section{CGM gas opacity estimates from scaling relations}
\label{appendix2}

We summarise here the scaling relations used in Section \ref{subsec:line_opacities} to obtain an estimate of the expected gas opacities in the CGM for the various species considered in this paper. We make the hypothesis that the scattering medium is made of circumgalactic gas enclosed in DM haloes and thus we resort to virial quantities to compute the opacity of hydrogen and metal ions.  Here, we assume a flat $\Lambda$CDM Universe with a matter density parameter $\Omega_{\rm m,0}=0.3$, baryonic fraction $f_{\rm b}=0.15$ and Hubble constant $H_{\rm 0}=70$ km s$^{-1}$ Mpc$^{-1}$.\\

The scaling relations between virial mass, radius, temperature and velocity are given by:
\begin{equation}
\label{app_eq1}
 \dfrac{M_{\rm vir}}{\frac{4}{3}\pi R_{\rm vir}^3} = \bar{\rho}_{\rm h}  \: \mbox{ , } \\  
 V_{\rm vir} = \sqrt{\dfrac{GM_{\rm vir}}{R_{\rm vir}}}\: \mbox{ , } \\ 
 T_{\rm vir} = \dfrac{\mu m_{\rm p}}{2k_{\rm b}} V_{\rm vir}^2
\end{equation}
where $G$ is the gravitational constant, $k_{\rm b}$ is the Boltzmann constant and $m_{\rm p}$ is the proton mass. The mean molecular weight, $\mu$, usually takes values between 0.5 and 2 depending on the chemical composition and ionisation state of the gas, so here we simply assume $\mu=1$. The average mass density of a DM halo, $\bar{\rho}_{\rm h}$, can also be expressed as: 
\begin{equation}
\label{app_eq2}
   \bar{\rho}_{\rm h} = \Delta_{\rm vir} \Omega_{\rm m,0} H^2_{\rm 0}(1+z)^3
\end{equation}
where the virial overdensity parameter is defined as $\Delta_{\rm vir}=(18\pi^2+82x-39x^3)/(x+1)$ \citep[][with $x=\Omega_{\rm m}(z)-1$]{Bryan_1998}. $\Delta_{\rm vir}$ is a (weak) decreasing function of redshift, converging to $\approx 18 \pi^2$ at $z\gtrsim 2$. 

Assuming a uniform gas density profile and ignoring wind kinematics, we can express the integrated line centre opacity of the wind enclosed within the virial radius of a halo as:\\[-8pt]
\begin{equation}
\label{app_eq3}
\tau_{\rm g} = \sigma_{\rm 0} n_{\rm g} R_{\rm vir}
\end{equation}
Here, $n_{\rm g}$ is the average (number) density of the gas that we write as:
\begin{equation}
\label{app_eq4}
n_{\rm g} = f_{\rm b} f_{\rm CGM} \dfrac{\bar{\rho}_{\rm h}}{\mu m_{\rm p}}
\end{equation}
where $f_{\rm CGM}$ is a free parameter corresponding to the fraction of the baryonic mass of the halo contained in the CGM. Here we arbitrarily set it to 0.25. 
The cross-section at line centre is given by :
\begin{equation}
\label{app_eq5}
\sigma_{\rm 0}=\frac{\sqrt{\pi} e^2 f_{lu} c}{m_e \nu_{lu} b}
\end{equation}
Assuming virialised gas at constant temperature within the halo, the gas velocity dispersion, $b$, can be approximated from the virial temperature, or equivalently, the virial velocity, according to Eq. \ref{app_eq1}.\\

Combining Eq. \ref{app_eq4} and \ref{app_eq5} to rewrite Eq. \ref{app_eq3} for the gas opacity, the dependencies on virial quantities basically cancel out, such that $\tau_{\rm g}$ only depends on atomic physics parameters and redshift in our simplified formalism :
\begin{equation}
\label{app_eq6}
\tau_{\rm g} = \frac{3 e^2 f_{lu}}{4 \sqrt{\pi} G \mu m_e m_{\rm p} \nu_{lu}} f_{\rm b} f_{\rm CGM} \Bigl(\frac{1}{2}\Delta_{\rm vir} \Omega_{\rm m,0}\Bigl)^{1/2} H_{\rm 0}(1+z)^{3/2} 
\end{equation}

For a primordial hydrogen abundance ratio, the \lya opacity as a function of the \hi neutral fraction is then $\tau_{Ly\alpha} = 0.76 f_{\rm HI} \tau_{\rm g}$. For solar metal abundance relative to hydrogen $(X/H)_{\odot}$, the opacity for metal ions can be similarly written as a function of the fraction of singly-ionised species, i.e. $\tau_{X} = 0.76 f_{\rm X^{+}} (X/H)_{\odot} \tau_{\rm g}$.

In this model, we deliberately made simplifying hypotheses to get rough estimates of the line opacities. We note that assuming different values for the CGM gas fraction ($f_{\rm CGM}$), the CGM extent ($R_{\rm vir}$), the metal abundance ratio $(X/H)$, or accounting for the depletion of metals onto dust would linearly rescale $\tau_{Ly\alpha}$ and $\tau_{X}$. While metal abundance ratios can easily span a relatively wide range (e.g. $(X/H) \approx 0.1-2 (X/H)_{\odot}$) across the population of SF galaxies depending on their mass, recent star formation history and/or redshift, dust depletion is also likely to vary by (at least) a factor of a few as a function of gas metallicity \citep{Roman_Duval_2022}. The CGM gas fraction is challenging to measure observationally but existing estimates suggest that $f_{\rm CGM}$ is comprised between $\approx 5-60\%$ according to various simulations and observations \citep{Werk_2014,Tumlinson_2017,Khrykin_2024}, bracketing our adopted fiducial value of $25\%$. Furthermore, including wind kinematics would considerably reduce the \lya opacity since the its associated cross-section is significantly lower in the wing than in the core (see Figure \ref{fig:cross}). The impact of winds on metal line opacities is likely less important because, in this case, interactions occur predominantly close to the line centre through continuum pumping of photons on the blue side of the resonance regardless of the wind velocity (see spectra shown in Section \ref{sec:integrated_spec}).

\label{lastpage}

\end{document}